\documentclass[prd,twocolumn,showpacs,amsmath,amssymb,showkeys,showpacs,superscriptaddress]{revtex4}

\usepackage{ifthen}
\usepackage{graphicx}
\usepackage{dcolumn}
\usepackage{bm}
\usepackage{color}
\usepackage{hyperref}

\newcommand{\Mpc}{\ \text{Mpc}}
\newcommand{\hMpc}{\ h^{-1}\text{Mpc}}
\newcommand{\ihMpc}{\ h\text{Mpc}^{-1}}
\newcommand{\hkpc}{\ h^{-1}\text{kpc}}

\newcommand{\hMs}{\ h^{-1} M_\odot}
\newcommand{\eh}[1]{\exp{\left[#1\right]}}
\newcommand{\tim}[1]{\times 10^{#1}}
\newcommand{\unit}[1]{\ \text{#1}}
\newcommand{\derivd}{\mathrm{d}} 

\newcommand{\tr}[1]{#1}

\newcommand{\ba}{\begin{eqnarray}}
\newcommand{\ea}{\end{eqnarray}}

\newcommand{\be}{\begin{equation}}
\newcommand{\ee}{\end{equation}}

\def\nn{\nonumber}

\renewcommand{\vec}[1]{\bm{#1}}
\allowdisplaybreaks
\begin{document}

\preprint{}

\title{An algorithm for the direct reconstruction of the dark matter correlation
function\\ from weak lensing and galaxy clustering}
\author{Tobias Baldauf}
 \email{baldauf@physik.uzh.ch}
\affiliation{Institute for Theoretical Physics, University of Zurich, Zurich, Switzerland}%
\author{Robert E. Smith}%
\affiliation{Institute for Theoretical Physics, University of Zurich, Zurich, Switzerland}%
\author{Uro\v{s} Seljak}
\affiliation{Institute for Theoretical Physics, University of Zurich, Zurich, Switzerland}
\affiliation{Physics Department, Astronomy Department and Lawrence Berkeley National Laboratory, University of California, Berkeley, CA, USA}
\affiliation{Ewha University, Seoul, South Korea}
\author{Rachel Mandelbaum}%
\affiliation{Department of Astrophysical Sciences, Princeton University, Peyton  
Hall, Princeton, NJ, USA}%

\date{\today}
\begin{abstract}
The clustering of matter on cosmological scales is an essential probe
for studying the physical origin and composition of our Universe.
To date, most of the direct studies have focused on shear-shear weak lensing 
correlations, but it is also possible to extract the dark matter clustering 
by combining galaxy-clustering and galaxy-galaxy-lensing measurements.
In order to extract the required information, one must relate the
observable galaxy distribution to the underlying dark matter distribution. In
this study we develop in detail a method that can constrain the dark matter
correlation function from galaxy clustering and galaxy-galaxy-lensing
measurements, by focusing on the correlation coefficient between the galaxy and
matter overdensity fields. Our goal is to develop an estimator that maximally
correlates the two. To generate a mock galaxy catalogue for testing purposes, we
use the Halo Occupation Distribution approach applied to a large ensemble of
$N$-body simulations to model pre-existing SDSS Luminous Red Galaxy sample
observations. Using this mock catalogue, we show that a direct comparison
between the excess surface mass density measured by lensing and its
corresponding galaxy clustering quantity is not optimal. We develop a new
statistic that suppresses the small-scale contributions to these observations
and show that this new statistic leads to a cross-correlation coefficient that
is within a few percent of unity down to $5 \hMpc$. Furthermore, the residual
incoherence between the galaxy and matter fields can be explained using a
theoretical model for scale-dependent galaxy bias, giving us a final estimator
that is unbiased to within 1\%, so that we can reconstruct the dark matter
clustering power spectrum at this accuracy up to $ k \sim 1 \ihMpc$. We also
perform a comprehensive study of other physical effects that can affect 
the analysis, such as redshift space distortions and differences in radial
windows between galaxy clustering and weak lensing observations. We apply the
method to a range of cosmological models and explicitly show the viability of
our new statistic to distinguish between cosmological models.
\end{abstract}


\maketitle
\section{Introduction}

The current paradigm for the history of our Universe, also known as the
$\Lambda$CDM cosmology, comes along with dark ingredients that have not yet been
directly detected: Cold Dark Matter (hereafter CDM) and Dark Energy
\cite{Komatsu2008}. CDM particles constitute about 20\% of the total energy
budget of the Universe, and whilst there is no confirmed direct detection of
them in a laboratory experiment, the indirect astrophysical evidence supporting
their existence is substantial. However, even more puzzling is the existence and
true physical nature of Dark Energy, which contributes roughly 75\% of the
total energy budget of the Universe and is responsible for driving the late-time
accelerated expansion of spacetime.

The dark matter power spectrum and its real space equivalent, the correlation
function, contain a wealth of cosmological information, e.\,g.\ on neutrino
mass, dark energy equation of state and the initial conditions of the Universe.
Thus it is a key goal of cosmology to infer these quantities from observables.
However, to achieve this requires a solid understanding of the galaxy bias --
the relation between the observable galaxies and the underlying dark matter
density field. This understanding is especially important for the interpretation
of ongoing and upcoming surveys, such as
SDSS\footnote{\texttt{http://www.sdss.org}},
DES\footnote{\texttt{http://www.darkenergysurvey.org}},
PanSTARRS\footnote{\texttt{http://pan-starrs.ifa.hawaii.edu/public/}} and EUCLID
\cite{Refregier2008}.

The reconstruction of the CDM distribution is usually based on the assumption,
that galaxies trace the matter density field, i.\,e.\ that on large scales the
galaxy density field equals the matter density field times a parameter
known as the bias. The resulting galaxy correlation function can then
be expressed as
\begin{equation}
	\xi_\text{gg}(r)=b^2 \xi_\text{mm}(r),
\end{equation}
and similarly for the power spectrum in $k$-space. The subtlety in the
standard approach is that the bias has to be determined empirically,
leading to uncertainties in the amplitude of the matter correlation,
which finally complicates studies of the rate of change of matter fluctuations
with time (the growth factor). Furthermore there is evidence for a non-trivial
scale dependence of galaxy bias \cite{Cole2005,Smith2007,Sanchez2008}. Hence it
is of great importance to devise methods that allow a direct reconstruction of
the dark matter correlation function from observables. One of the most promising
observational probes of dark matter on cosmological scales is the gravitational
lensing.

We will focus our attention on a specific weak lensing technique, halo-galaxy
lensing, which involves measurement of the shape distortions around foreground
dark matter haloes in which galaxies form. Often the foreground object (lens)
will be an individual galaxy, in which case this technique is called
galaxy-galaxy lensing, but it can also be applied to groups and clusters. 
Since the first attempts to detect galaxy-galaxy lensing by \cite{Tyson1984},
the quality of the data has been improved vastly by deeper and wider surveys.
Halo-galaxy lensing has now been measured with relatively high signal-to-noise
and as a function of a wide variety of properties of the lens galaxies, groups
and clusters \cite{Guzik2002,Hoekstra2003,Seljak2005,Mandelbaum2006} . It has
become clear in these studies that galaxy-galaxy lensing contains much
information about the mass distribution around galaxies, and has the potential
to measure dark matter halo radii, shapes, concentrations and masses
\cite{Mandelbaum2006b,Mandelbaum2008,Evans2009,Okabe2009} as well as the
distribution of matter within the Universe
\cite{McKay2001,Sheldon2004,Schneider2005,Sheldon2009a,Sheldon2009b}.

The interpretation of the signal in terms of the link between galaxies
and dark matter is, however, complicated by the fact that (except for
galaxy clusters) galaxy-galaxy lensing is only detectable by stacking
the signal from many lenses. Theoretical modelling of the galaxy-galaxy
lensing has been done both with numerical simulations
\cite{Hayashi2007,Hilbert2009} and with the halo model
\cite{Guzik2001,Mandelbaum2005}. The combination of lensing and clustering seems
to hold the potential to put constraints on cosmological parameters
\cite{Yoo2006,Cacciato2008}.

In this paper our main objective is to develop a method that recovers
a statistic closely related to the matter correlation function
from a joint analysis of lensing and clustering observations. This
method is presented together with a theoretical motivation and tests on
simulated galaxy samples.  

The starting point for these simulated galaxy samples are cosmological $N$-body
simulations, which are a standard tool to investigate
the non-linear evolution of the CDM density field. Despite their
statistical power for describing the large scale structure of the
Universe, pure dark matter simulations have the disadvantage that one
must supplement them with a prescription for galaxy formation in order
to reproduce the surveyed galaxy distributions. We work in the
standard paradigm of hierarchical galaxy formation: {\em galaxies
  only form in dark matter haloes} \cite{WhiteRees1978}. Hence, the
problem is reduced to that of relating galaxies to dark matter haloes,
and we do this using the Halo Model approach and in particular the
Halo Occupation Distribution (for a review see \cite{Cooray2002}). Here we are
focused on obtaining mock galaxy
catalogues for the Luminous Red Galaxies (LRGs), a subset of galaxies
observed with the Sloan Digital Sky Survey (SDSS). Our modelling builds
on earlier approaches by \cite{Zheng2008,Reid2008}.

The paper breaks down as follows: in \S \ref{sec:obs} we review the
basics of weak gravitational lensing, an important probe of the dark
matter on cosmological scales. Then in in \S \ref{sec:crosscorr} we introduce
our main analysis tool, the cross-correlation coefficient.  Theoretical
modelling of the latter is carried
out in \S \ref{sec:theo}. In \S \ref{sec:num} we describe the simulations and
the mock galaxy catalogues that we use to test our new method. The
results of the numerical studies on the cross-correlation coefficient and the
reconstructed matter statistic are discussed in \S \ref{sec:results}. The effect of redshift space distortions and radial window functions on the observational implementation of our method are
explored in \S \ref{sec:rs}. \S \ref{sec:varcosm} is devoted to the
cosmology-dependence of our results. Finally, in \S \ref{sec:discuss} we will summarise and
discuss our findings.

\section{Observables} \label{sec:obs}

\subsection{Halo-Galaxy Lensing} 

Weak gravitational lensing is one of the main probes for the dark
matter distribution in the Universe (see
\cite{Bartelmann2001,Schneider2006,Schneider2006a} for reviews). In this study
we focus on a specific weak lensing technique known as
halo-galaxy or galaxy-galaxy lensing. In this technique, one infers the
tangential shear $\gamma_\text{t}$ around foreground objects from the
deformation of background galaxy images. Since the shear is weak, one
must average over a large number of background galaxies to obtain good
signal to noise. The estimated $\gamma_\text{t}$ can then be related
to the projected mass distribution around the foreground lens
galaxies. The key quantity is the differential excess surface mass
density \cite{SquiresKaiser1996,Schneider2006a},
\begin{equation}
\Delta\Sigma_\text{gm}(R)=\overline
\Sigma_\text{gm}(R)-\Sigma_\text{gm}(R)=
\Sigma_\text{crit}\left\langle\gamma_\text{t}(R,\varphi)\right\rangle_{\varphi},\label{eq:deltasigmagamma}
\end{equation}
where $\Sigma_{\rm gm}$ is the projected surface mass density,
$R\approx \theta D_\text{l}$ is the comoving transverse distance between lens
and source galaxies with angular separation $\theta$, and subscripts g and
m refer to galaxies and mass, respectively \footnote{We will use a
lowercase $r$ to denote 3D radii, whereas the uppercase $R$ is
reserved for 2D radii in the projected statistics. Furthermore all
distances are comoving and expressed in terms of $\hMpc$.}. In the
above equation we also introduced the comoving angular diameter distance  to the
lens galaxy $D_\text{l}$ and the mean surface mass density within a
circular aperture,
\begin{equation}
	\overline\Sigma_{\rm gm}(R)=\frac{2}{R^2}\int_0^R\Sigma_{\rm
gm}(R')R'\,\derivd R'\ .
\end{equation}
The critical surface mass density
\begin{equation}
 \Sigma_\text{crit}=\frac{c^2}{4\pi G} \frac{D_\text{s}}{D_\text{ls}D_\text{l}
}
\end{equation} 
is a geometrical factor with
$D_\text{s},D_\text{l},D_\text{ls}$ being the angular diameter
distances to the source, the lens and between lens and source,
respectively. Galaxy-galaxy lensing stacks the signal of large numbers
of foreground and background galaxies and thus $\Sigma_\text{crit}$
has to be understood as an effective quantity for the lens and source
distribution. It is sensitive to the
cosmological model, including the matter density parameter
$\Omega_\text{m}$.

Since the deflections are measured around foreground galaxies, the mass
profile is directly related to the galaxy-matter cross-correlation
function
\begin{equation}
	\Sigma_\text{gm}(R)= \Omega_\text{m}\rho_\text{crit}
 \int_{-\infty}^{+\infty} g_\text{l}(\chi) 
\left[1+\xi_\text{gm}(\sqrt{R^2+\chi^2})\right]\derivd\chi\ ,\label{eq:sigmaint}
\end{equation}
with integration along the line of sight $\chi$. The critical
density is defined as $\rho_\text{crit}(a)=3 H^2(a)/8\pi G$, where
$H(a)\equiv \dot{a}/a$ is the Hubble parameter.  Here we include the
radial window function $g_\text{l}(\chi)$ (see e.\,g.\ \cite{Guzik2001}) that
describes the dependence of lensing strength on the distribution of the lens
mass and depends on the lens and source positions. Note that the additional
constant $1$ in the integrand drops out on computing $\Delta\Sigma_\text{gm}(R)$
with Eq.~\eqref{eq:deltasigmagamma}.

In principle the excess surface mass density $\Delta\Sigma(R)$ could be
integrated to yield the projected galaxy-matter correlation function $w(R)$,
which in turn can be deprojected to $\xi(r)$ using an Abel formula. Lensing
observations are, however, subject to noise, that is amplified when
reconstructing the correlation function $\xi_\text{gm}(r)$. Consequently, we try
to minimise the manipulations on the data, and rather transform theoretical
predictions accordingly.


\subsection{Projected correlation functions}

In addition to the mass distribution around galaxies one may also
observe the distribution of galaxies themselves. A convenient way to
quantify the clustering between the tracer fields A and B is the
projected correlation function \cite{DavisPeebles1983},
\begin{equation}	
	w_\text{AB}(R)=\int_{-\infty}^{+\infty} g_{\rm g}(\chi)
	\xi_\text{AB}\bigl(\sqrt{\chi^2+R^2}\bigr)\derivd\chi\label{eq:wab},
\end{equation}
where $g_{\rm g}(\chi)$ is a window function and where for instance we
are interested in:
$\text{AB}=\left\{\text{gg},\text{gm},\text{mm}\right\}$. The line of
sight integration partially removes redshift space distortions, which
are an issue in the three dimensional correlation function $\xi(r)$
(see \S \ref{sec:rs} for a discussion of the residual effects).  Based
on the projected galaxy clustering $w_\text{gg}(R)$, we now define two
statistics that correspond more closely to the lensing observable
$\gamma_t$:
\ba
\Delta\Sigma_{\rm gg}(R) & \equiv & 
\rho_\text{crit} \left[\overline w_\text{gg}(R)-w_\text{gg}(R)\right]\ ; \\
\Delta\Sigma_{\rm mm}(R) & \equiv & \Omega_\text{m}^2\rho_\text{crit} 
\left[\overline w_\text{mm}(R)-w_\text{mm}(R)\right]\ ,
\ea
In these equations, we have multiplied by the critical density in order to
achieve the same dimensions as $\Delta\Sigma_\text{gm}$. The prefactor
$\Omega_\text{m}^2$ in the definition of $\Delta\Sigma_\text{mm}$ accounts for
the fact that it is a two-point statistic of matter density. Lensing is
sensitive to the total density of matter $\rho_\text{m}$ (which is proportional
to $\Omega_\text{m}$), while for galaxy clustering we usually remove the
dependence on the mean density of galaxies and work only with the 
density contrasts $\delta_\text{g}=(\rho_\text{g}-\overline
{\rho_\text{g}})/\overline{\rho_\text{g}}$.

So far we have not specified the window functions for the line of
sight integrations in Eqs.~\eqref{eq:sigmaint} and \eqref{eq:wab}, $g_{l}(\chi)$ and $g_{\rm g}(\chi)$. In galaxy-galaxy lensing the inhomogeneous mass distribution between the observer and the source contributes to the final distortion. Consequently the window
for lensing is typically very broad and is fixed by the geometrical
setup of the source-lens-observer system, i.\,e.\ the radial
distribution of the lens and source samples. For galaxy clustering
studies, when provided with accurate redshifts, the window function can be
constructed straightforwardly and we shall assume a narrow top-hat
around the lens positions. We take the thickness of the top-hat to be
$\Delta \chi\approx 100 \hMpc$, which is a compromise between adding
uncorrelated noise and increasing the signal.

To simplify our investigations further we measure $\{\Delta\Sigma_{\rm
gg}(R),\Delta\Sigma_{\rm gm}(R),\Delta\Sigma_{\rm mm}(R)\}$ from our
simulations with top-hat window functions of the same length. The
estimates are obtained in real space and we quantify the effects of
window functions, integration lengths and redshift space distortions
on the result separately in \S \ref{sec:rs}. This approach enables us to disentangle the
intrinsic properties of the mass and tracer fields and the systematic
effects induced by the measurement technique.

Note that since our main goal is to develop an algorithm for
reconstructing the mass clustering, 
we have also assumed that the correlation function
is estimated over a region of space where the galaxy selection
function does not vary significantly, hence one must be careful when 
applying it to the galaxies close to the edge of the survey. 


\section{Cross-Correlation Coefficient} \label{sec:crosscorr}

The cross-correlation coefficient between two density fields A and B
may be defined using the correlation function $\xi$ as \footnote{We
  will follow the convention of denoting the cross-correlation
  coefficient with a lowercase $r_\text{cc}$, and add the subscript cc
  in order to avoid confusion with the radius $r$.}
\begin{equation}
	r_\text{cc,AB}^{(\xi)}(r)=\frac{\xi_\text{AB}(r)}{\sqrt{\xi_\text{AA}(r)\xi_\text{BB}(r)}} \ ,
\label{eq:ccxi}
\end{equation}
and is a measure of the statistical coherence of the two
fields \cite{Tegmark1998,Pen1998,Dekel1999,Seljak2004,Bonoli2008}. If
$r_\text{cc}=1$ then the fields are fully correlated and there exists a
deterministic mapping between the fields. This behaviour would be expected for
any scale-dependent, deterministic, linear bias model of haloes or galaxies:
$\xi_\text{gm}(r)=b(r)\xi_\text{mm}(r)$,
$\xi_\text{gg}(r)=b^2(r)\xi_\text{mm}(r)$.  On the other hand, if
$r_\text{cc}\ne 1$ then the fields are incoherent, and for the local model of
galaxy formation, this may arise due to stochasticity and non-linearity in the
bias relation \cite{Dekel1999}. The $r_\text{cc}$ constructed from real-space
statistics can be $>1$ (unlike in Fourier space), since $\xi_\text{gg}$ has the
shot noise subtracted off; this behaviour will be seen in several places in this
work.

Studying the cross-correlation coefficient $r_\text{cc,hm}^{(\xi)}$ of the
haloes in the numerical simulations used for this work we find that
the cross-correlation coefficient of haloes is close to unity on large
scales and decreases below unity on scales below $10 \hMpc$ similarly
for a large range of halo masses $1.3 \tim{13} \hMs \leq M \leq 3
\tim{15} \hMs$ \cite{Smith2009,Seljak2009}.

As was already mentioned, it is a key goal of cosmology to recover the dark
matter correlation function from the observations. In this context it
is important to quantify how well galaxies trace the underlying dark
matter density field, which inspires us to examine the
cross-correlation coefficient between the matter and galaxy fields. 
One approach is to measure the excess
surface mass density from galaxy-galaxy lensing using Eq.~\eqref{eq:sigmaint}. 
In this case, we define cross-correlation coefficient by replacing the
correlation functions in Eq.~\eqref{eq:ccxi} with the corresponding
excess surface mass densities,
\begin{equation}
	r_\text{cc,gm}^{(\Delta\Sigma)}(R)=\frac{\Delta\Sigma_\text{gm}(R)}{\sqrt
	{\Delta\Sigma_\text{gg}(R)\Delta\Sigma_\text{mm}(R)}}.
\end{equation}
Due to our definition of $\Delta\Sigma_\text{mm}$ and
$\Delta\Sigma_\text{gg}$ the prefactors $\Omega_\text{m}$ cancel and the
resulting statistic is only dependent on the ratio of the correlation functions.
Thus $r_\text{cc,gm}^{(\Delta\Sigma)}(R)$ is expected to approach unity on
linear scales.

The excess surface mass density $\Delta\Sigma(R)$ measures the
difference between the surface mass density averaged over an aperture of
radius $R$ and the actual value at the boundary of the
aperture. Consequently it combines information from small scales, which 
are highly non-linear and stochastic, and larger, linear scales, where stochasticity
is believed to be small. To remove part of the
incoherence introduced by the non-linear clustering process we
introduce a new statistic $\Upsilon(R)$, that we call the Annular
Differential Surface Density (hereafter ADSD). This statistic
eliminates the contributions to $\Delta\Sigma(R)$ from
small scales as follows:
\ba
\Upsilon(R;R_0) & \equiv & \Delta\Sigma(R)-\frac{R_0^2}{R^2}\Delta\Sigma(R_0) \ ; \\
            &   =    & \frac{2}{R^2}\int_{R_0}^{R}\derivd R' R' \Sigma(R') \nn  \\
	    &   & - \frac{1}{R^2}\Bigl[R^2 \Sigma(R)-R_0^2\Sigma(R_0)\Bigr]\label{eq:upsilon}.
\label{eq:Upsilon}
\ea
Setting the cutoff radius to $R_0=0$ the new statistic $\Upsilon$
reduces to $\Delta\Sigma$. Note that $\Upsilon$ is completely
independent of the correlation function on scales below $R_0$. Our
motivation in subtracting out small-scale contributions was to recover
a statistic that does not mix small and large scales. \tr{Thus we
  suggest the choice $R_0\approx 2 r_\text{vir}$, where $r_\text{vir}$ is the average virial 
radius of the host haloes of the galaxy sample under consideration. On scales
below two virial radii, the intra-halo non-linear clustering dominates,
whereas the weakly non-linear scales exceeding $2 r_\text{vir}$ can be modelled 
by simulations and perturbation theory. We suggest a conservative choice of 
$R_0$ to avoid problems in the transition region between small and large scales 
even if the signal-to-noise ratio is slightly degraded.
The virial radii have to be inferred from a mass estimator such as X-ray or 
gravitational lensing. The latter has the advantage that the same observation
can be used to infer the mass and the ADSD statistic. In a companion paper
\cite{Mandelbaum2009} we show that the ADSD, with a cutoff radius
$R_0\approx 0.25 r_\text{vir}$, 
can also be used to avoid statistical and systematical uncertainties about the 
inner parts of the halo profiles and thus is a viable tool to calculate cluster 
masses using an iterative procedure.}

One may calculate the cross-correlation coefficient of the ADSD
\be
	r_\text{cc}^{(\Upsilon)}(R)=\frac{\Upsilon_\text{gm}(R)}{\sqrt
	{\Upsilon_\text{gg}(R)\Upsilon_\text{mm}(R)}}.
	\label{eq:ccgam}
\ee
%
%
\begin{figure}[!t]
	\centering{
	\includegraphics[width=0.49\textwidth]{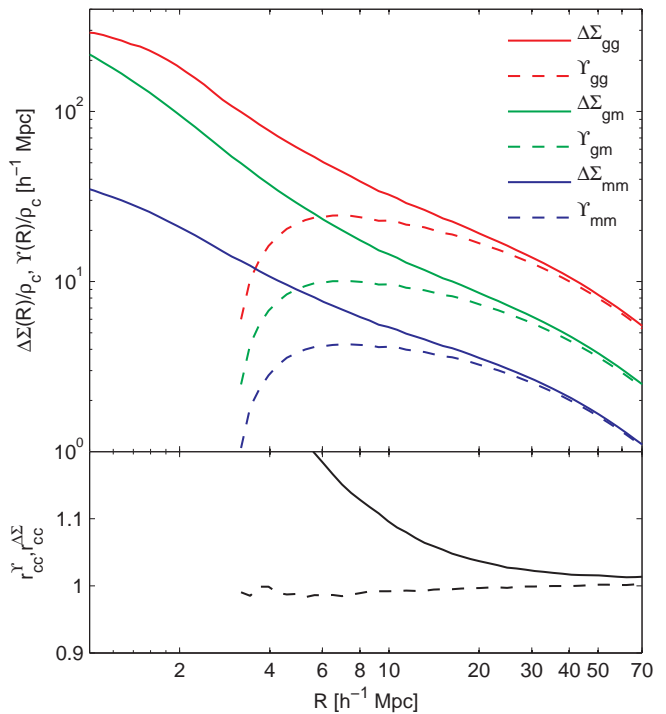}}
	\caption{\emph{Top panel: }Excess surface mass density $\Delta\Sigma(R)$
(solid)
          and ADSD $\Upsilon(R;R_0)$
          (dashed) with $R_0=3 \hMpc$ for our fiducial cosmological
          model and the luminosity-threshold LRG sample. We show the
          statistics for the galaxy auto-correlation (top red), the
          galaxy-matter cross-correlation (central green) and the
          matter auto-correlation (bottom blue). The upturn of the
          cross-correlation towards small scales leads to a
          cross-correlation coefficient in excess of unity as we will
          see later. \emph{Bottom panel: }Cross-correlation coefficient of the
clustering statistics shown in the top panel. The bare excess surface mass
density (solid) leads to strong deviations from unity, whereas the ADSD with
$R_0=3\hMpc$ (dashed) recovers a cross-correlation close to unity.}
	\label{fig:sigmaupsilon}	
\end{figure}
%
%
In Fig.~\ref{fig:sigmaupsilon}, we plot both the excess surface mass
density and the ADSD $\Upsilon$ defined from the galaxy
auto-correlation, matter auto-correlation and their
cross-correlation. As galaxies we choose a model for Luminous Red Galaxies (LRGs), as discussed in 
more detail in \S \ref{sec:num}. We observe that the galaxy-galaxy and
galaxy-matter
excess surface mass densities are not
multiples of the matter correlation function on small scales, so that we expect
a cross-correlation different from unity for the bare statistic. This result is
seen in the bottom panel of Fig.~\ref{fig:sigmaupsilon}, 
where the deviations from unity extend to scales above $10 \hMpc$. 
Subtracting the signal at $R_0=3 \hMpc$ as in Eq.~\ref{eq:upsilon} to get
$\Upsilon$, we remove these non-linearities and recover similar shapes for all
three functions. As a result, the cross-correlation coefficient is now much
closer to unity on all scales above $R_0$, as seen in the bottom of
Fig.~\ref{fig:sigmaupsilon}.

\begin{figure*}
	\centering
	\includegraphics[width=1.0\textwidth]{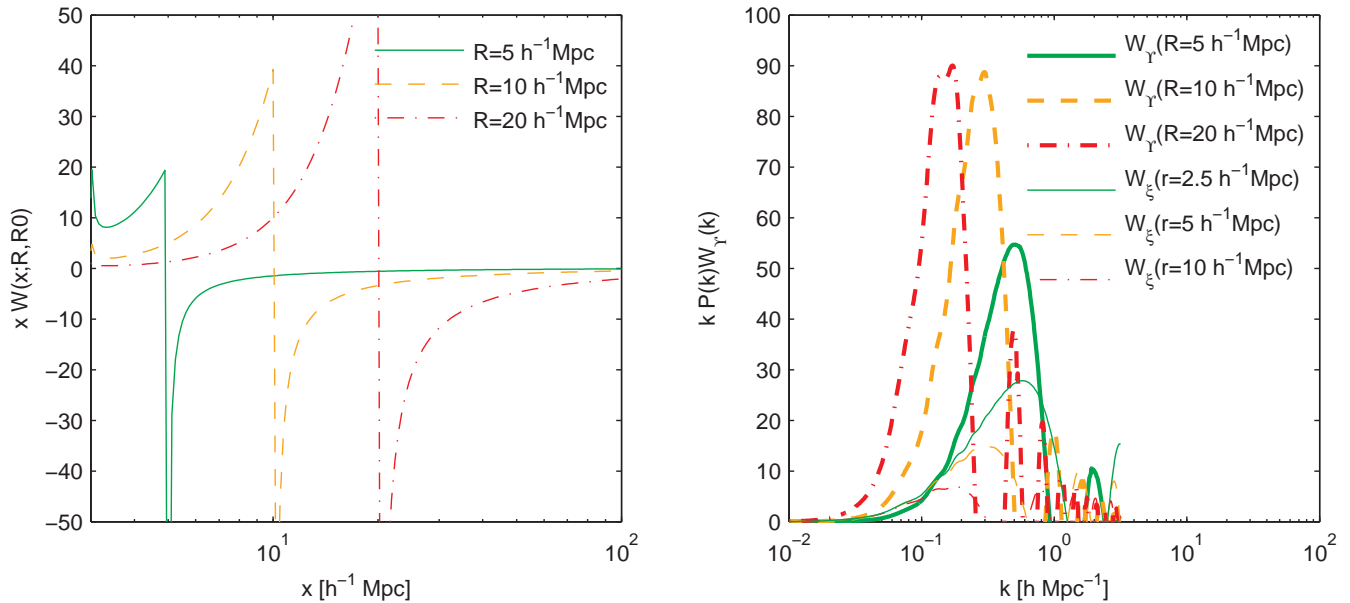}
	\caption{\emph{Left panel: }Window functions for
          $\Upsilon(R;R_0)$ with $R_0=3 \hMpc$ in real space. We show
          the window function for $R=5,10,20 \hMpc$ as green solid,
          orange dashed and red dash-dotted lines
          respectively. \emph{Right panel: }Window functions for
          $\Upsilon(R;R_0)$ with $R_0=3 \hMpc$ in $k$-space. We show
          the products of window function and power spectrum for
          $R=5,10,20 \hMpc$ as green solid, orange dashed and red
          dash-dotted line respectively. For reference we also plot the
window for the correlation function as thin lines. Note that we multiplied with
          $x$ and $k$ respectively to account for the logarithmic
          scale on the ordinate axis.}
	\label{fig:wWindow}
	\label{fig:PWindow}
\end{figure*}

Both the projected correlation function and the ADSD are defined by
integrals of the correlation function weighted by a kernel. The
projected correlation function can be written as
\be
w(R)= \int_0^{+\infty}\xi(x)W_w(x;R)x\,\derivd\ln{x}\ ,
\ee
where the window function is written,
\be
W_w(x)=
\frac{2x}{\sqrt{x^2-R^2}}\Theta(x-R)\Theta(\sqrt{\chi_\text{max}^2+R^2}-x)\ .
\ee
Here $\Theta(x)$ is the Heaviside step function. Thus $w(R)$ has
contributions only from scales $x\geq R$, and the window function is peaked at
$x=R$.

The ADSD defined in Eq.~\eqref{eq:upsilon} involves a radial average
and subtraction of $w(R)$. Both operations can be included in the
integration kernel, and we may write the ADSD as
\be
 \frac{\Upsilon(R;R_0)}{\rho_\text{crit}}= \int_0^{+\infty}\xi(x) 
W_\Upsilon(x;R,R_0)x\,\derivd\ln{x}\ .
\label{eq:Upsreal}
\ee
where the window function for $\Upsilon(R,R_0)$ is written,
\begin{widetext}
\be
W_\Upsilon(x;R,R_0)=\frac{4x}{R^2}\left[\sqrt{x^2-R_0^2}\Theta(x-R_0)-
\sqrt{x^2-R^2}\Theta(x-R)\right]-\frac{2x}{R^2}\left[\frac{R^2
    \Theta(x-R)}{\sqrt{x^2-R^2}}-\frac{R_0^2
    \Theta(x-R_0)}{\sqrt{x^2-R_0^2}}\right]\ .
\label{eq:upswindow}
\ee
\end{widetext}
The scale dependence of the integration kernels reveals the scales in
the correlation function that are dominating.

In Fig.~\ref{fig:wWindow} we show the window function
$W_\Upsilon(x;R,R_0)$ for three different radii and $R_0=3
\hMpc$. Since $\xi(x)$ approximately follows a decreasing power-law,
the leading contribution is at the scale $R$, where the sign
changes. This sign-change is due to the subtraction
$\Delta\Sigma_\text{AB}=\rho \left[\overline w_\text{AB}(R)-w_\text{AB}(R)
  \right]$ and is the same as for $\Delta \Sigma$.
For $\Delta \Sigma$ the window is exactly compensated, meaning that it
integrates 
to zero, hence $\Delta \Sigma$ is insensitive to adding a mean density
component, the 
so called mass sheet degeneracy. This compensation is fortunate, since it
means that this statistic is insensitive to the long wavelength modes that can
move $w(R)$ up and down, i.\,e.\ the long wavelength sampling variance affects
$w(R)$ on all scales. The compensated window also makes $\Delta \Sigma$ 
less sensitive to the redshift space distortions, as discussed in \S
\ref{sec:rs}. The statistic $\Upsilon$, though not exactly compensated, retains
most of these beneficial properties, while at the same time eliminating small
scale clustering information. 

The scales probed by the statistic $\Upsilon$ are however more obvious
in the power spectrum. The conversion from $P(k)\to \Upsilon(R;R_0)$
can be written as
\be
 \Upsilon(R;R_0)=\int P(k) k W_\Upsilon(k;R,R_0)\,\derivd\ln{k},\label{eq:pwin}
\ee
where the window function is no longer given by a simple analytical
form due to the spherical Bessel functions occurring in the Fourier
transform.  In Figure \ref{fig:PWindow} we show this window function
multiplied with the power spectrum. From this plot we see that for a
cutoff radius $R_0=3 \hMpc$, $\Upsilon$ essentially probes scales down
to $k \approx 1\ h\unit{Mpc}^{-1}$. The integrand in \eqref{eq:pwin} is
peaked at the scale $k\approx \pi/R$, and is strongly oscillatory on
small scales.

For later use we plot the window function for the correlation function
$W_\xi(k;r)=k^2 \sin{kr}/{kr}$ that relates $P(k)$ and $\xi(r)$ via
$\xi(r)=V/(2\pi)^3 \int  k P(k) W_\xi(k;r)\,\derivd \ln{k}$.



\section{Theoretical Modelling of scale dependent bias} \label{sec:theo}


\begin{figure*}[p]
	\centering
	\includegraphics[width=1.0\textwidth]{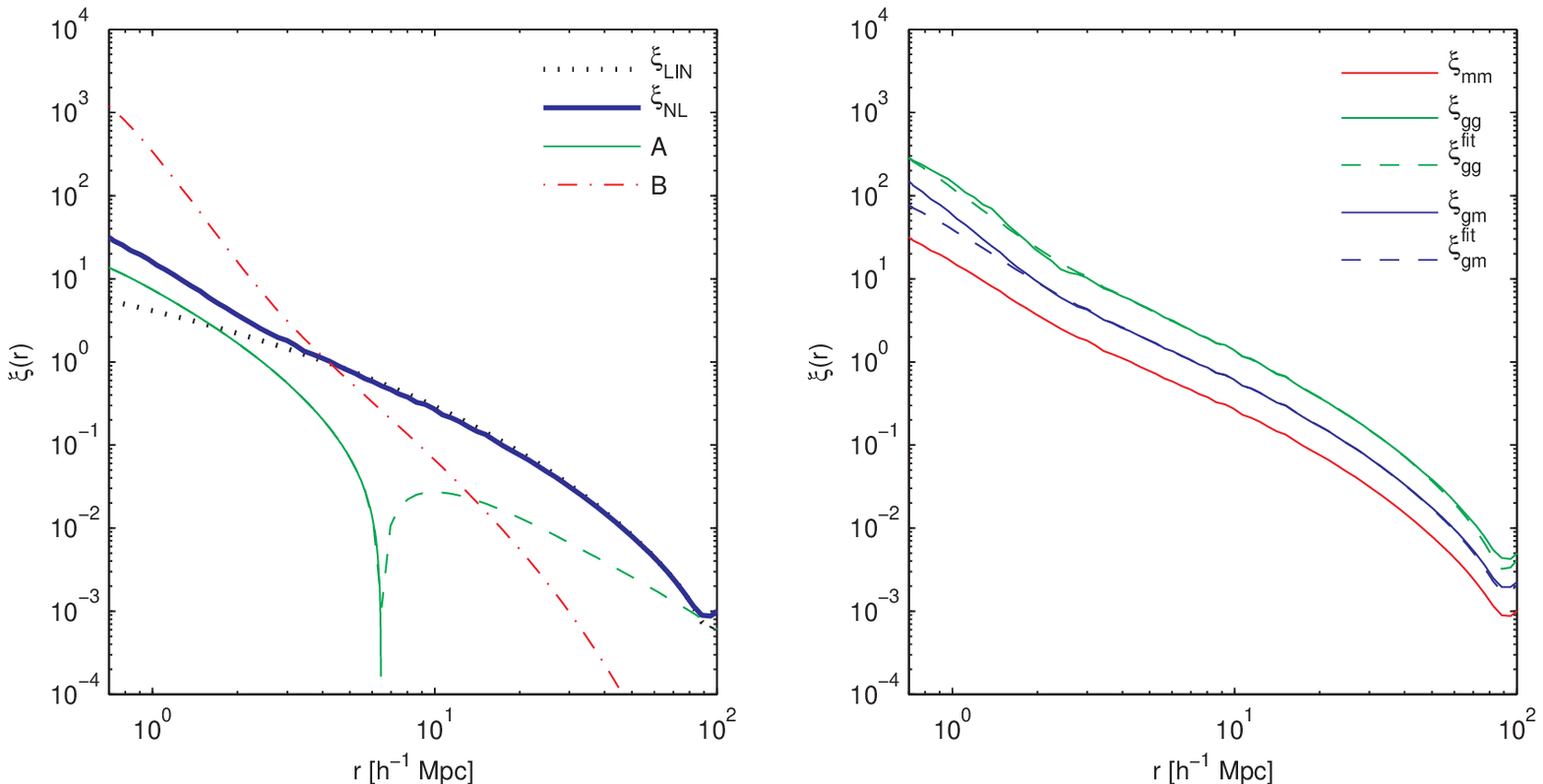}
	\caption{\emph{Left panel: }Non-linear corrections to the real space
	correlation function as function of radial separation
	calculated for redshift $z=0.23$. We show the linear (black dotted) and
non-linear matter correlation function
	(thick blue solid) as well as the $B$ (red dash-dotted) and
	$A$ (green solid and dashed) correction terms. The dashed
	portion of the graph of $A(r)$ denotes the range where it is
	negative. \emph{Right panel: }Perturbation theory fit (dashed) over
scales $6 \hMpc \leq r \leq 80 \hMpc$
	to the measured galaxy-correlation functions (solid) of the
	luminosity-threshold sample. The fits to $\xi_\text{gg}$
	(upper green), $\xi_\text{gm}$ (central blue) and a joint fit
	provide consistent results.  We are not expecting a good agreement on
scales below $r\approx 3\hMpc$, where the correlation function is dominated by
non-linear clustering.}
	\label{fig:mcdonaldxi}
	\vspace{1cm}
	\includegraphics[width=1.0\textwidth]{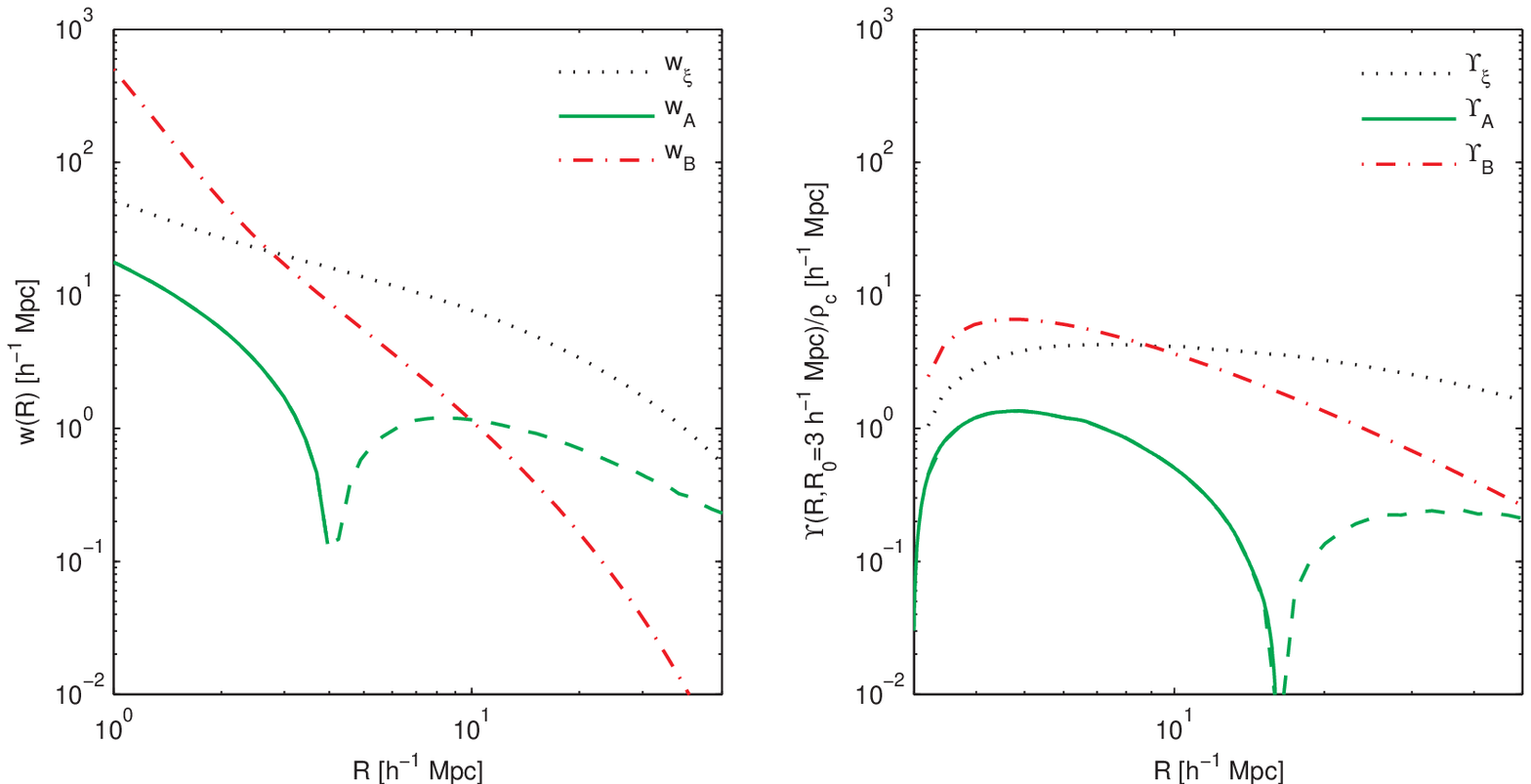}
	\caption{\emph{Left panel: }Perturbation terms for the
	projected surface mass density $w$. We show the non-linear
	matter correlation function (black dotted) as well as the
	$w_\text{B}$ (red dash-dotted) and $w_\text{A}$ (green dashed
	and solid) terms. Note that the dashed part of the latter
	curve is negative.  \emph{Right panel: } Perturbation terms
	for the ADSD and $R_0=3 \hMpc$. Again we show the non-linear
	matter statistic (black dotted) together with the
	$\Upsilon_\text{A}$ (green dashed and solid) and
	$\Upsilon_\text{B}$ (red dash-dotted) terms. Note that the
	scale at which the $\Upsilon_\text{B}$ term becomes comparable
	to the non-linear correlation is shifted further out to
	$R\approx 9\hMpc$.}
	\label{fig:mcdonaldwy}
\end{figure*}



In this section we will use cosmological perturbation theory
(for a review see \cite{Bernardeau2002}) to predict the
cross-correlation coefficient. Our discussion is based on a Taylor
expansion of the galaxy density field in terms of the matter
overdensity $\delta$
\begin{equation}
\rho_\text{g}=\rho_0+\rho_0' \delta+\frac{1}{2}\rho_0''
\delta^2+\frac{1}{6}\rho_0''' \delta^3+\epsilon+\mathcal{O}(\delta^4)
\end{equation}
Such an expansion is only valid on scales exceeding the virial radius
of dark matter haloes, since it contains no mechanism of halo
exclusion or the radial distribution of galaxies within their host
halo. Absorbing potentially divergent terms into the bias and shot
noise parameters, \cite{Smith2007,McDonald2006} showed that the auto-
and cross-power spectrum of a biased tracer field can be written up to
fourth order in the matter density field as
\ba
	P_\text{gm}(k)&=&b_1 P_\text{NL}(k)  + b_2 A(k),\label{eq:mcdpt1}\\
	P_\text{gg}(k)&=&b_{1}^2 P_\text{NL}(k)  + 2 b_{1}b_2 A(k)+
\frac{b_{2}^2}{2} B(k)+N\label{eq:mcdpt2},
\ea
where $N$ is the renormalized shot noise, $b_1$ and $b_2$ are the
renormalized bias parameters and $P_\text{NL}$ is the non-linear power
spectrum. The calculation of the latter can be carried out with
any perturbative technique, e.\,g. standard perturbation theory
\cite{Bernardeau2002}, renormalized perturbation theory
\cite{Crocce2006,Crocce2006b} or Lagrangian perturbation theory. The
advantage of this renormalisation of the bias parameters is that there
is no artificial smoothing scale involved in the above
expansion. However this comes at a price: the bias and shot
noise are no longer given \emph{ab-initio} from theory, but have to be
determined empirically.

The correction terms $A(k)$ and $B(k)$ introduced in the above
equation are defined as
\ba
A(k)\!\!&=&\!\!\int\frac{\derivd^3q}{(2\pi)^3}P_\text{lin}(q)P_\text{lin}(|\vec
k -\vec q|)
F_2(\vec q, \vec k-\vec q),\\
B(k)\!\!&=&\!\!\int\frac{\derivd^3q}{(2\pi)^3}P_\text{lin}(|\vec q|)
\left[P_\text{lin}(|\vec k -\vec q|)-P_\text{lin}(q)\right],
\ea
where
\begin{equation}
F_2(\vec k_1, \vec k_2)=\frac{5}{7}+
\frac{1}{2}\frac{\vec k_1 \cdot \vec k_2}{k_1 k_2}\left(\frac{k_1}{k_2}+
\frac{k_2}{k_1}\right)+\frac{2}{7}\left(\frac{\vec k_1 \cdot \vec k_2}{k_1
k_2}\right)^2
\end{equation}
is the second order standard mode coupling kernel.  In the above
integrals, one can use the linear power spectrum, because the integrals
are already fourth order in the matter density field $\delta$.  \par
Due to the linearity of the expressions in Eqs.~\eqref{eq:mcdpt1} and
\eqref{eq:mcdpt2} the transformation to real space is straightforward
\begin{align}
	\xi_\text{gm}(r)=&b_1 \xi_\text{NL}(r)  + b_2 A(r)\ ,\\
	\xi_\text{gg}(r)=&b_{1}^2 \xi_\text{NL}(r)  +2 b_{1}b_2
A(r)+\frac{b_{2}^2}{2} B(r) \,
\end{align}
with $\xi_\text{mm}(r)=\xi(r)$, and $A(r)$, $B(r)$ being the Fourier
transforms of $A(k)$, $B(k)$ respectively. It is easy to show that
$B(r)=\xi^2-\sigma^2 \delta^\text{D}(\vec r)$, where $\sigma$ is the variance
of the power spectrum.
Figure~\ref{fig:mcdonaldxi} shows the terms contributing to the galaxy
auto- and cross-correlation functions as well as a fit to the
correlation functions measured in our numerical simulations. The
$A(r)$ term is positive on small scales and changes sign at $r\approx
6 \hMpc$. The $B(r)$ term affecting the auto-correlation dominates
over the matter correlation function on scales below $r\lesssim 4
\hMpc$.

Let us for later convenience define the parameter combination
\begin{equation}
\alpha\equiv \frac{b_2}{b_1},
\end{equation}
As shown by \cite{Seljak2009} in the
regime where $A(r)\ll \xi(r)$ and $B(r)\ll \xi(r)$ the
cross-correlation coefficient can be written as:
\begin{align}
r_\text{cc}^{(\xi)}=&\frac{\xi+\alpha A}{\sqrt{\xi(\xi+2\alpha A +
\alpha^2B/2)}}\ ;\\
\approx &
1-\frac{1}{4}\alpha^2\frac{B}{\xi}-\alpha^2\frac{A^2}{\xi^2}-\frac{1}{4}
\alpha^3\frac{AB}{\xi^2}\ ;\\
	\approx& 1-\frac{1}{4}\alpha^2\frac{B(r)}{\xi(r)}\ ;\\
	\approx& 1-\frac{1}{4}\alpha^2\xi(r)\ .\label{eq:ccxitheo}
\end{align}
Obviously the model predicts a scale dependent cross-correlation that
is below unity on small scales and asymptotically approaches unity for
increasing $r$. As shown in \cite{Seljak2009}, the shape of the
cross-correlation coefficient of haloes measured in the simulations is
well described by the functional form of the above equation, and the
prefactor $\alpha$ is a weak function of halo mass. Clearly this
simple theoretical model is not able to cover the non-linear behaviour
inside the virial radius after shell-crossing. Qualitatively similar
predictions were presented by \cite{Matsubara1999} for the peak model
of \cite{BBKS1986}.

We will now proceed to develop the results repeated here for the
readers' convenience for use in our investigations.  A result
similar to Eq.~\eqref{eq:ccxitheo} remains valid if we
consider the projected correlation function, since the integration
along the line of sight is a linear operation. We therefore have:
\begin{align}
	w(R)=&\int_{-\chi_\text{max}}^{+\chi_\text{max}}
\xi\bigl(\sqrt{R^2+\chi^2}\bigr)\derivd\chi\ ;\\
	w_A(R)=&\int_{-\chi_\text{max}}^{+\chi_\text{max}}
A\bigl(\sqrt{R^2+\chi^2}\bigr)\derivd\chi\ ;\\
	w_B(R)=&\int_{-\chi_\text{max}}^{+\chi_\text{max}}
B\bigl(\sqrt{R^2+\chi^2}\bigr)\derivd\chi\ .
\end{align}
Furthermore, the manipulations that lead to the excess surface mass
density, or more generally to the ADSD
$\xi(r) \to \Delta\Sigma(R) \to \Upsilon(R)$ are linear in the fields
$A$, $B$ and $\xi$. Consequently the corresponding terms have the same
order as their underlying statistic and we can write,
\begin{equation}	
	r_\text{cc}^{(\Upsilon)}(R)=1-\frac{1}{4}\alpha^2\frac{\Upsilon_\text{B}
(R)}{\Upsilon_\text{mm}(R)}\ .
	\label{eq:ccgamma}
\end{equation}
Note that to evaluate the term $\Upsilon_\text{B}(R)$ we only need to replace
$\xi(r)$ with $\xi^2(r)$ 
in Eq.~\eqref{eq:Upsreal}. 
The effective value of $\alpha=\left\langle b_2 \right\rangle/\left\langle b_1
\right\rangle$ for our galaxy catalogues can be estimated using the mean bias
parameters from the peak-background-split \cite{Sheth1999,Scoccimarro2001},
\begin{equation}
  \left\langle b_i \right\rangle=\frac{\int n(M)\left\langle N(M) \right\rangle
b_i(M) dM}{\int n(M)\left\langle N(M) 
    \right\rangle \derivd M} \ \ i=1,2,
\end{equation}
where $n(M)$ is the halo mass-function and $\left\langle
N(M)\right\rangle$ is the halo occupation number. For the rest of this
work we will adopt $\alpha=0.26$, which is close to peak-background split
predictions 
\cite{Seljak2009}.  An alternative
approach, accounting for the renormalised nature of the parameters,
would be to fit for the model parameters by matching theoretical and
measured correlation functions as shown in the right panel of
Figure~\ref{fig:mcdonaldxi}. This second approach provides results
that are consistent with the peak-background-split result.

In Figure~\ref{fig:mcdonaldwy} we plot the correction terms
contributing to the projected correlation function and the ADSD. The
prerequisite for the Taylor expansion to be applicable is that the
correction terms $w_\text{A}/w_\text{mm}\ll1$ and
$w_\text{B}/w_\text{mm}\ll1$ or
$\Upsilon_\text{A}/\Upsilon_\text{mm}\ll1$ and
$\Upsilon_\text{B}/\Upsilon_\text{mm}\ll1$. We see that these
assumptions are violated below $R=3 \hMpc$ for the projected correlation
function and below $R=9 \hMpc$ for the ADSD.  For the non-linear
correlation function $\xi$ we use the matter correlation
function measured in the simulations.


\section{Numerical Modelling}\label{sec:num}

With modern large supercomputers and well developed algorithms it is
now possible to model the evolution of the dark matter density
field on cosmological scales reasonably well. However, what one
observes are not dark matter haloes but galaxies. Supplementing the distribution
of dark matter in a simulation box with the
galaxy distribution corresponding to a particular galaxy sample would
in principle require an understanding of the process of
galaxy formation. An \emph{ab initio} treatment of all the baryonic
processes is difficult, and requires, e.\,g.\ full treatment of the
hydrodynamics, atomic and radiative heating and cooling of gas at high
resolution. Owing to the large computational cost, state of the art
simulations are restricted to relatively small scales and lack
sufficient volume to extract statistically relevant information on
cosmological scales \cite{Agertzetal2009}. Thus we pursue a
statistical approach to populate the haloes identified in a suite of
large-scale $N$-body simulations with galaxies.


\subsection{The Simulations}

Our numerical results are based on the Z\"{u}rich horizon
``\texttt{zHORIZON}'' simulations, a suite of $30$ pure
dissipationless dark matter simulations of the $\Lambda$CDM cosmology
in which the matter density field is sampled by $N_p=750^3$ dark
matter particles. The box length of $1500 \hMpc$, together with the
cosmological parameters given in Table \ref{tab:cosmoparam}, then implies
a particle mass of $M_\text{dm}=5.55 \times 10^{11} \hMs$. This
simulation volume enables high precision studies of the fluctuations
in the $\Lambda$CDM model on scales up to a few hundred comoving
megaparsecs \cite{Smith2008a}.

The simulations were carried out on the \texttt{ZBOX2} and
\texttt{ZBOX3} computer-clusters of the Institute for Theoretical
Physics at the University of Zurich using the publicly available
\texttt{GADGET-II} code \cite{Springel2005}. 
The force softening length of the simulations used for
this work was set to $60\hkpc$, consequently limiting our
considerations to larger scales. The transfer function at redshift $z=0$ was
calculated using the \texttt{CMBFAST} code of \cite{Seljak1996} and then
rescaled to the initial redshift $z_i=50$ using the linear growth factor. For
each simulation, a realisation of the power
spectrum and the corresponding gravitational potential were
calculated. Particles were then placed on a Cartesian grid of spacing
$\Delta x=2 \hMpc$ and displaced according to a second order
Lagrangian perturbation theory. The displacements and initial
conditions were computed with the \texttt{2LPT} code of
\cite{Scoccimarro1998,Crocceetal2006}.

The cosmological parameters for the simulations were inspired by the
best fit values released by the WMAP3 analysis of the cosmic microwave
background \cite{Spergel2003,Spergel2007}, and can be taken from Table
\ref{tab:cosmoparam}. Throughout the paper we adopt this parameter set as our
fiducial cosmological model.

Our effective volume is $V=27 \ h^{-3}\text{Gpc}^3$. 
For each of the simulation outputs, gravitationally bound structures
were identified using the \texttt{B-FoF} algorithm kindly provided by
Volker Springel. The linking length in this Friends-of-Friends halo
finder was set to $0.2$ of the mean inter-particle spacings, and haloes
with less than $20$ particles were rejected. All together we resolve
haloes with $M>1.2 \times 10^{13} \hMs$.
\begin{table}[ht]
	\centering
	\begin{tabular}{llllllll}
	\hline \hline
	&$\Omega_\text{m}$&$\Omega_\Lambda$&$h$&$\sigma_8$&$n_s$&$w$&$N_\text{e}
$\\
	\hline FID&$0.25$ &$0.75$&$0.7$&$0.8$&$1.0$&$-1$&8\\ C1&$0.25$
	&$0.75$&$0.7$&$0.8$&$0.95$&$-1$&4\\ C2&$0.25$
	&$0.75$&$0.7$&$0.9$&$1.0$&$-1$&4\\ C3&$0.2$
	&$0.8$&$0.7$&$0.8$&$1.0$&$-1$&4\\ C4&$0.3$
	&$0.7$&$0.7$&$0.8$&$1.0$&$-1$&4\\ \hline \hline
	\end{tabular}
	\caption{Cosmological parameters adopted for our
	investigations. Matter density parameter, dark energy density
	parameter, dimensionless Hubble parameter $H_0= 100 h
	\unit{km}\unit{s}^{-1}\unit{Mpc}^{-1}$, power spectrum
	normalisation, primordial power spectrum slope, dark energy
	equation of state $p=\omega \rho$, number of simulation
	outputs. The first line is our fiducial model. In order to
	evaluate the cosmology dependence of our results we use four
	other cosmologies denoted as C1-C4.}
	\label{tab:cosmoparam}
\end{table}


\subsection{HOD Modelling I - Luminosity Threshold Sample}

The statistical model used to populate the haloes with galaxies is
known as the Halo Occupation Distribution (HOD), which is closely related to the
Halo Model of Large Scale Structure (for a review see
\cite{Cooray2002}). The HOD assumes that galaxies form in the dark
matter potential wells, because only there can baryons cool with
sufficient efficiency. To translate this idea into a
quantitative model, one must fix the following ingredients:
\begin{enumerate}
	\item Number of galaxies that occupy a halo of mass $M$
	\item Radial distribution of galaxies within the halo
\end{enumerate}
Theories of galaxy formation suggest a division into central and
satellite galaxies. Central galaxies are those that reside at the
minimum of the potential well for host dark matter haloes. In contrast,
satellite galaxies orbit the central galaxy and are presumed to have
their own associated subhalo within the larger host halo. Furthermore we will
assume that the number of satellite galaxies is a function of host halo mass
only and neglect any environmental influences. The basic assumption of the model
is that bright galaxies will not be able to live in low mass haloes, since
there is not enough cold gas to form such galaxies. Equivalently, the halo
mass can be represented by the virial radius $r_\text{vir}$, defined by the
condition that the density within $r_\text{vir}$ equals 200 times the critical
density $\rho_\text{crit}$.

Let us start by considering the HOD required to model a
luminosity-threshold sample of galaxies, and later in
\S\ref{ssec:HODbin} we will describe the necessary adaptations required
for the more complex luminosity bin sample.

First, we decide whether a halo of given mass contains a
central galaxy at its potential minimum. It is reasonable to
assume that a threshold in galaxy luminosity corresponds to a
threshold in halo mass, but in practice it is necessary to take
into account the scatter in the luminosity-mass relationship. We do
this by appropriately smoothing the mass threshold. Following
\cite{Reid2008,Padmanabhan2008}, we take the mean number of central
galaxies occupying a mass $M$ halo to be:
\begin{equation}
	\left\langle N_\text{cen}\right\rangle=
\text{erfc}\left[-\frac{\ln\left(M/M_\text{cut}\right)}{\sqrt{2}\sigma}\right],
\end{equation}
where $\text{erfc}(x)=1-\text{erf}(x)$ is the complementary error
function, and $M_{\rm cut}$ and $\sigma$ are parameters to be
determined from the data. This relationship is then used as a sampling
probability for the Bernoulli distribution: operationally this amounts to
drawing a random number, $T\in[0,1]$, and if $T<\left<N_{\rm
cen}\right>$ then we place a central. This central galaxy is assumed
to be formed by the baryons cooling in the dark matter potential well
and subsequent collisions with satellite galaxies that approach the
halo centre due to dynamical friction.

Satellite galaxies that orbit the halo center mostly
originate from the merging of haloes already containing a central
galaxy. Subhalo counts in high resolution $N$-body dark matter
simulations have shown that the number of satellite galaxies follows a
Poisson distribution around an asymptotic power law
\cite{Kravtsov2004}. Hence we take,
\be
\left\langle N_\text{sat}\right\rangle_\text{c}(M)=
\begin{cases}
\left(\frac{M-M_\text{min}}{M_1}\right)^\alpha, & \text{if }
M>M_\text{min}\wedge N_\text{cen}\neq 0 \\
  0, & \text{otherwise}\ ,
\end{cases}
\ee
which introduces another three parameters to be determined: $M_{\rm
min}$, $M_1$ and $\alpha$. Finally, as a further constraint we impose
the condition that satellite galaxies can only reside in
haloes already containing a central galaxy.

The satellite galaxies are expected to be situated in the subhaloes
orbiting the halo centre. Our simulations lack sufficient resolution
to identify such dark matter substructures, so we instead sample the
galaxy positions from the dark matter particle positions. Compared to
a galaxy distribution following a profile this approach has several
advantages. Firstly, we avoid the assumption of a functional form for
the halo profile, instead profiting from the full triaxial dark matter
distribution. Secondly, we can assign the dark matter particle
velocities to the galaxies, which is useful for studies of the
redshift space distortions.

All together, we have five-dimensional parameter space
spanned by $\{M_\text{cut},\sigma,M_\text{min},M_1,\alpha\}$. We vary
these five parameters in order to generate galaxy catalogues that can
reproduce observed galaxy clustering and galaxy-galaxy lensing
measurements for the two LRG samples described in
\cite{Mandelbaum2005} and \S \ref{sec:refsamp} below.  For each point
in parameter space, we generate four galaxy catalogues per simulation
using different random seeds. We then calculate the average of the
clustering statistics of these four catalogues to remove some of the
stochasticity intrinsic to the HOD model, and finally compare to the
data. To reduce the dimension of the parameter space and thus the
computational costs, we use the observed abundance of the LRG sample to
impose a further constraint on the cutoff mass $M_\text{cut}$ by
demanding
\begin{equation}
	\overline{n}_\text{obs}=\int \derivd M n(M)\left\langle N_\text{tot}
	\right\rangle,\label{eq:numdens}
\end{equation}
where $n(M)$ is the halo mass-function \footnote{For the actual
calculation we use a spline fit to the mass-function measured from our
FoF halo catalogues, which is however well reproduced by the
\cite{Sheth1999} mass-function.} and the mean total number of galaxies
per halo is given by
\begin{equation}
	\left\langle N_\text{tot} \right\rangle=\left\langle
	N_\text{cen} \right\rangle\bigl[\left\langle N_\text{sat}
	\right\rangle+1\bigr],
\end{equation}
where the form of $\left\langle N_\text{cen} \right\rangle$ accounts for the
scatter in the luminosity-mass relationship.  Finally, we end up with
a 4-dimensional parameter space, which is sampled on a grid of points,
for each of which we calculate $M_\text{cut}$. 

\begin{figure}[!t]
	\centering
	\includegraphics[width=0.49\textwidth]{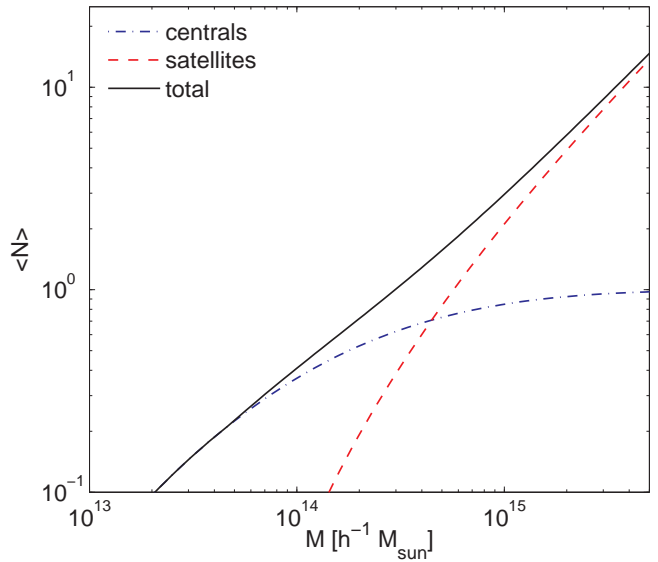}
	\caption{Mean galaxy number per halo as a function of halo
	mass for the luminosity-threshold sample and our best fit
	parameters from Table \ref{tab:bestfit}. We show the number of
	central galaxies (blue dash-dotted), satellite galaxies (red
	dashed) and the total number of galaxies (black solid). Note
	that the satellite number exceeds unity only for haloes with
	$M>6\tim{14} \hMs$, corresponding to virial radii exceeding
$r_\text{vir} \gtrsim 2 \hMpc$.}
	\label{fig:hod}
\end{figure}

Figure \eqref{fig:hod} shows the mass dependence of the central,
satellite and total halo occupation number for our best fit
luminosity-threshold galaxy samples. The total occupation number is
dominated by the central galaxies residing in the highly abundant low-mass
haloes. Satellite galaxies start to dominate only for masses of
$M\approx4\tim{14} \hMs$.


\subsection{HOD Modelling II - Luminosity Bin Sample}\label{ssec:HODbin}

To model a luminosity binned galaxy sample it is necessary to apply
some minor changes to our HOD modelling. Firstly, the halo mass of the
central galaxies will be a window rather than a threshold. Secondly,
we must drop the constraint that satellite galaxies live only in
haloes already hosting a central galaxy, because faint galaxies may orbit in
heavier haloes (that already host a central above the luminosity cutoff) as
satellites. For simplicity, we do not use the information about central LRGs
from the bright sample, but rather model the two samples
independently. Again the number of central galaxies is assumed to
follow a Bernoulli distribution, but with mean given by
\begin{equation}
	\left\langle N_\text{cen} \right\rangle=
\frac{1}{4}\text{erfc}\left[-\frac{\ln{M/M_{\text{cut},1}}}{\sqrt{2}\sigma}
\right]
\text{erfc}\left[\frac{\ln{M/M_{\text{cut},2}}}{\sqrt{2}\sigma}\right]\ ,
\end{equation}
where we have assumed that the central galaxy distribution is
symmetric in $\log M$ and that the mass-luminosity scatter is independent of
mass. This parametrisation introduces three free
parameters: $[M_\text{cut,1},M_\text{cut,2}]$ with a smoothing parameter
$\sigma$. One may of course conceive more complicated window
functions, however this approach introduces the least number of
additional free parameters into the modelling procedure whilst being
flexible enough to describe the data. 

For the satellite galaxy distribution, we again assume that the number
follows a Poisson distribution, with mean specified by
\begin{equation}
	\left\langle N_\text{sat}\right\rangle_\text{c}(M)=\begin{cases}
  \left(\frac{M-M_\text{min}}{M_1}\right)^\alpha,  & \text{if } M>M_\text{min}\\
  0, & \text{otherwise}\ .
\end{cases}
\end{equation}
Thus in total we must constrain six free parameters. However, we may
reduce the dimensionality of the problem by calculating the
appropriate lower mass cutoff $M_\text{cut,1}$ for each of the points
in the five-dimensional space spanned by
$\{M_\text{cut,2},\sigma,M_1,M_\text{min},\alpha\}$ according to
Eq.~\eqref{eq:numdens}.


\subsection{Reference Sample}\label{sec:refsamp}

In this study, we develop our analysis for application to the SDSS
spectroscopic LRG sample  \cite{York2000,Eisenstein2001}. The LRGs are typically
bright red ellipticals that are volume-limited within a much 
larger volume than the main galaxy sample of the SDSS. Thus they are
frequently used as an efficient tracer of large scale
structure. Furthermore, since the LRGs have been shown to live in the most
massive haloes of the Universe
\cite{Mandelbaum2006,Reid2008,Reid2008a,Zheng2008}, they can be effectively
probed with our $N$-body simulations.

The specific LRG samples that we tune our HODs to are presented in
\cite{Mandelbaum2006}. We model the galaxy-galaxy
lensing from that study, along with new projected correlation function
measurements of the same samples. In \cite{Mandelbaum2006}, the LRG samples were
split into two sub-samples, LRGbright and LRGfaint, based on the
$r$-band luminosities $k+e$-corrected to $z=0$. LRGbright is a
luminosity-threshold sample with a number density of
$\overline{n}=4\tim{-5}h^3\Mpc^{-3}$,
whereas LRGfaint is a luminosity-bin
sample with a number density of $\overline{n}=8\tim{-5}h^3\Mpc^{-3}$. As a
result, different strategies must be applied when modelling the sub-samples, as
discussed in the previous two subsections. The LRG sample under consideration
spans a redshift range $0.15 \leq z \leq 0.35$
with an effective redshift of $z_\text{eff}=0.24$. This effective redshift
was derived from the lensing analysis, since higher redshift lens galaxies are
downweighted by the lower number of source galaxies behind
them. As shown in \cite{Tegmark2006,Padmanabhan2008}, the clustering amplitude
of LRGs is independent of redshift due to a subtle balance between the redshift
evolutions of bias and growth. Therefore we  use the simulation outputs at
$z_\text{sim}=0.23$, very close to the effective lensing redshift, for our
numerical analysis.

\subsection{Fit results}

In Table \ref{tab:bestfit}, we quote the inferred HOD parameters for
the bright and faint samples when using the fiducial cosmological model. 
We decided to use the full covariance matrix for the fitting since there are
non-negligible correlations between $R$-bins, both in the
lensing and clustering measurements. The noise in the covariance matrix
increases the inferred $\chi^2$, an effect that has previously been investigated
by \cite{Hirata2004} and we think that theoretical covariance predictions could
improve the analysis. Imperfect modelling might of course also arise from the
fact that our simulation cosmology is not a perfect representation of the real
Universe.  Due to the computational costs per model and the high dimensionality
of the parameter space we have to restrict to coarse sampling of parameter
space. It is however not our goal to precisely constrain the HOD parameters, but
rather to obtain reasonable galaxy catalogues for the two LRG samples under
consideration and use them to test the $\Upsilon$ statistics. We compare to the
HOD parameters obtained by \cite{Reid2008}, who used an equivalent model,
finding a reasonable agreement, once we account for the fact that they
model the full LRG sample.

The galaxy catalogues have relatively low satellite fractions of
$\sim4.5\%$ for the threshold sample and $\sim10\%$ for the bin
sample. Fitting for the bias on linear scales ($18 \hMpc \leq r \leq 90 \hMpc$)
we obtain
$b=2.20\pm0.03$ and $b=1.97\pm0.03$ for the threshold and bin sample,
respectively.
%

 \begin{table}[h]
	\centering
	\begin{tabular}{lccccccc}
	\hline
	\hline
	 & $\overline n$ & $M_1$ & $M_\text{cut,l}$& $M_\text{cut,u}$ & $\alpha$
&
$\sigma$ &$M_\text{min}$ \\ 
	\hline
	LRGbright & $4.0$ & $40.0$& $17.8$ & --- & $1.05$ &$1.68$ & $4.6$ \\  
	LRGfaint & $8.0$ & $45.0$ & $5.0$ & $12.4$ & $0.40$ &$1.55$ & $5.7$\\
	\hline
	\hline
	\end{tabular}
	\caption{Best fit HOD parameters for the faint (f) and bright
	(b) sample: comoving number density, power law normalisation, low
	mass cutoff, high mass cutoff, power law exponent, smoothing and lower
	satellite cutoff. The number densities are in units of $1\tim{-5} \
	h^3\unit{Mpc}^{-3}$, and masses are in units of $10^{13}\hMs$.}
	\label{tab:bestfit}
 \end{table} 


\begin{figure*}[t]
	\centering
	\includegraphics[width=0.49\textwidth]{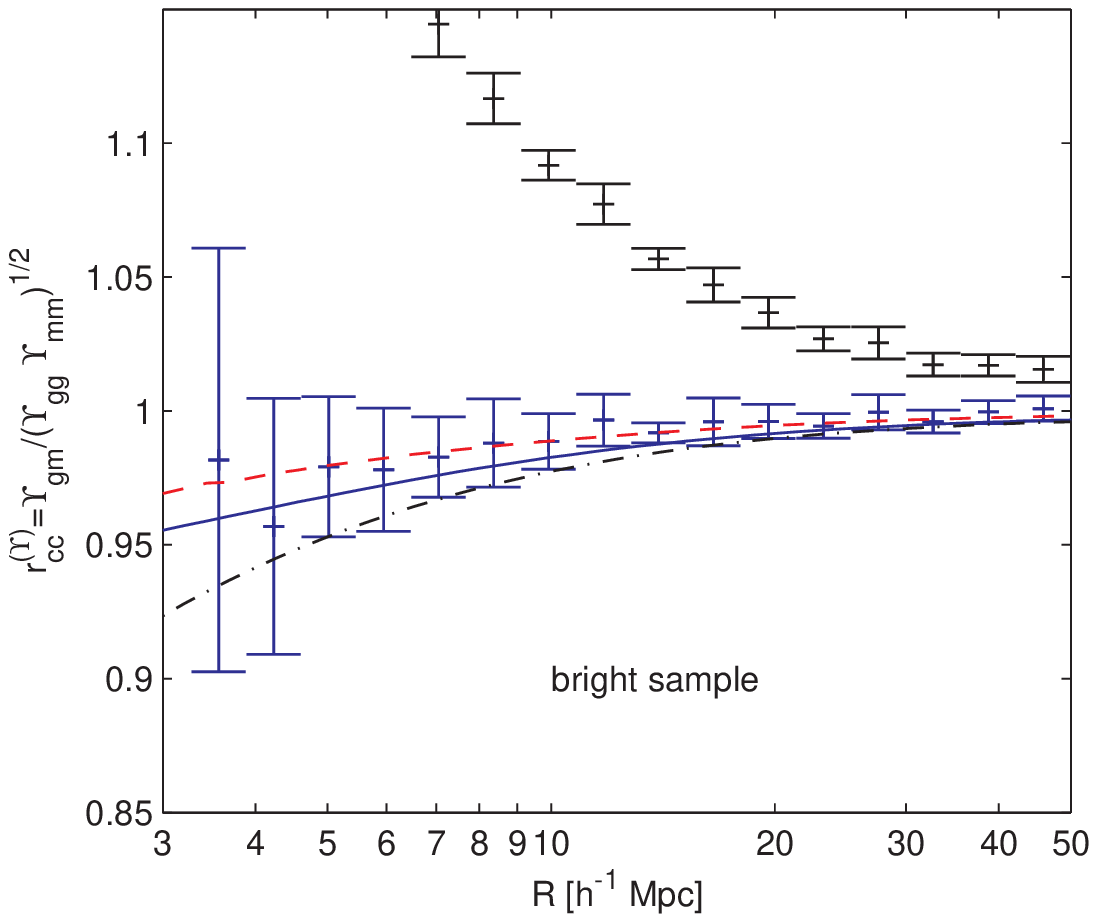}
	\includegraphics[width=0.49\textwidth]{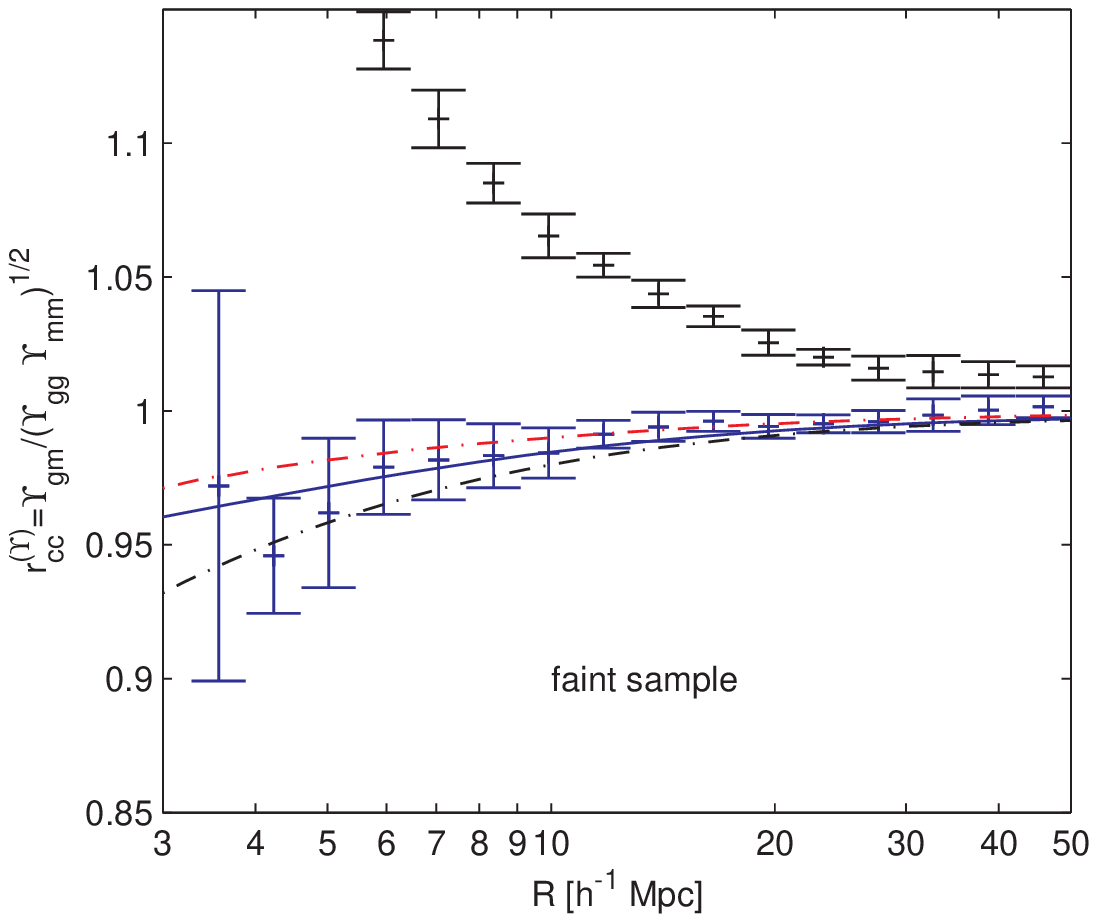}
	\caption{Cross-correlation coefficient of the ADSD $\Upsilon$
	for the luminosity-threshold (left panel) and luminosity bin
	sample (right panel). The trivial case $R_0=0$ (black with
	errorbars), corresponding to bare $\Delta\Sigma$, leads to a
	cross-correlation coefficient that is far from unity and
	furthermore strongly scale dependent. If we instead choose
	$R_0=3 \hMpc$ (blue with errorbars), inspired by the virial
	radii of the haloes under consideration, we restore a
	cross-correlation coefficient close to unity on the $4 \%$
	level for all scales $R>R_0$. Furthermore, we can model the residual
	scale dependence reasonably well using the perturbation theory
	expression of Eq.~\eqref{eq:ccgamma} (solid blue line), \tr{whereas
	the the bare $\Delta\Sigma$ deviates from the corresponding 
	perturbation theory result (black dash-dotted)}. Since the
	Taylor expansion is no longer justified for scales below $R=8
	\hMpc$, we also plot the full expression according to Equation
	\eqref{eq:mucorrfull} (red dashed) for $\Upsilon$.}
	\label{fig:rups}
	\vskip 10pt
	\includegraphics[width=0.49\textwidth]{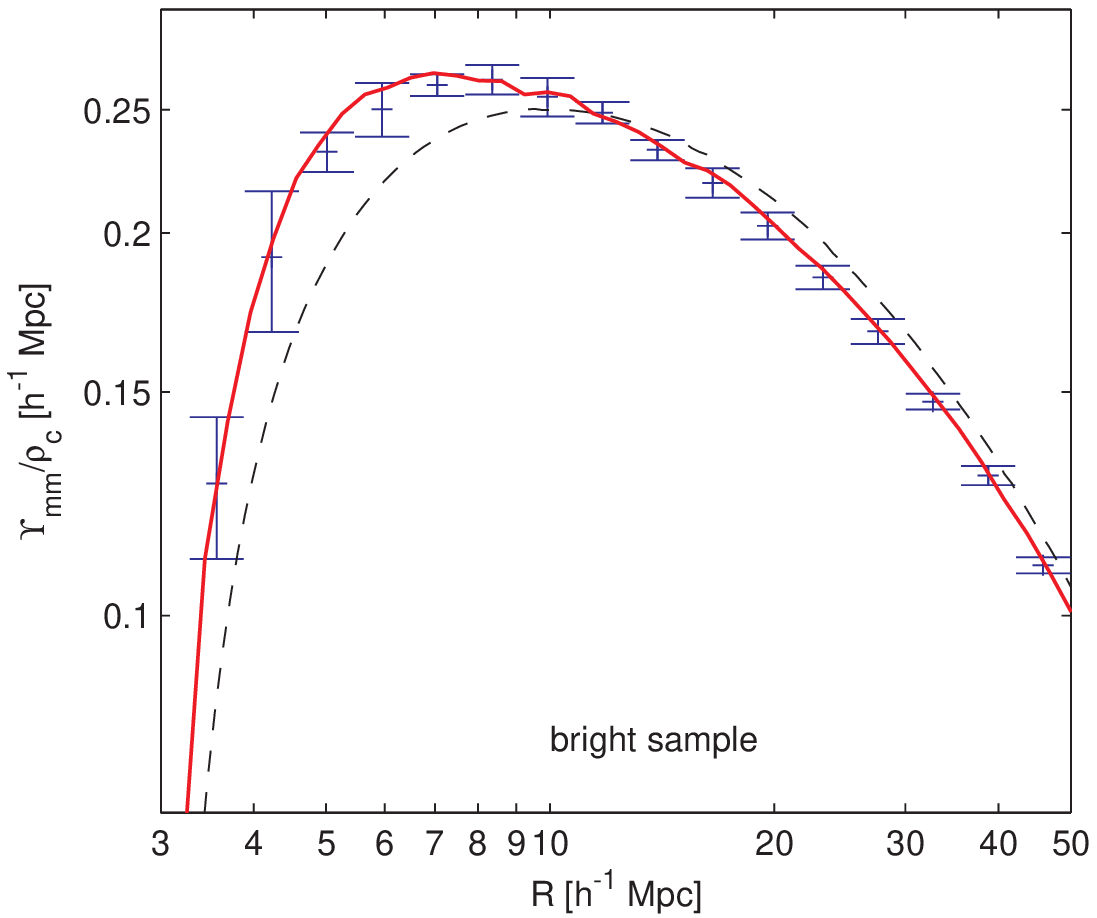}
	\includegraphics[width=0.49\textwidth]{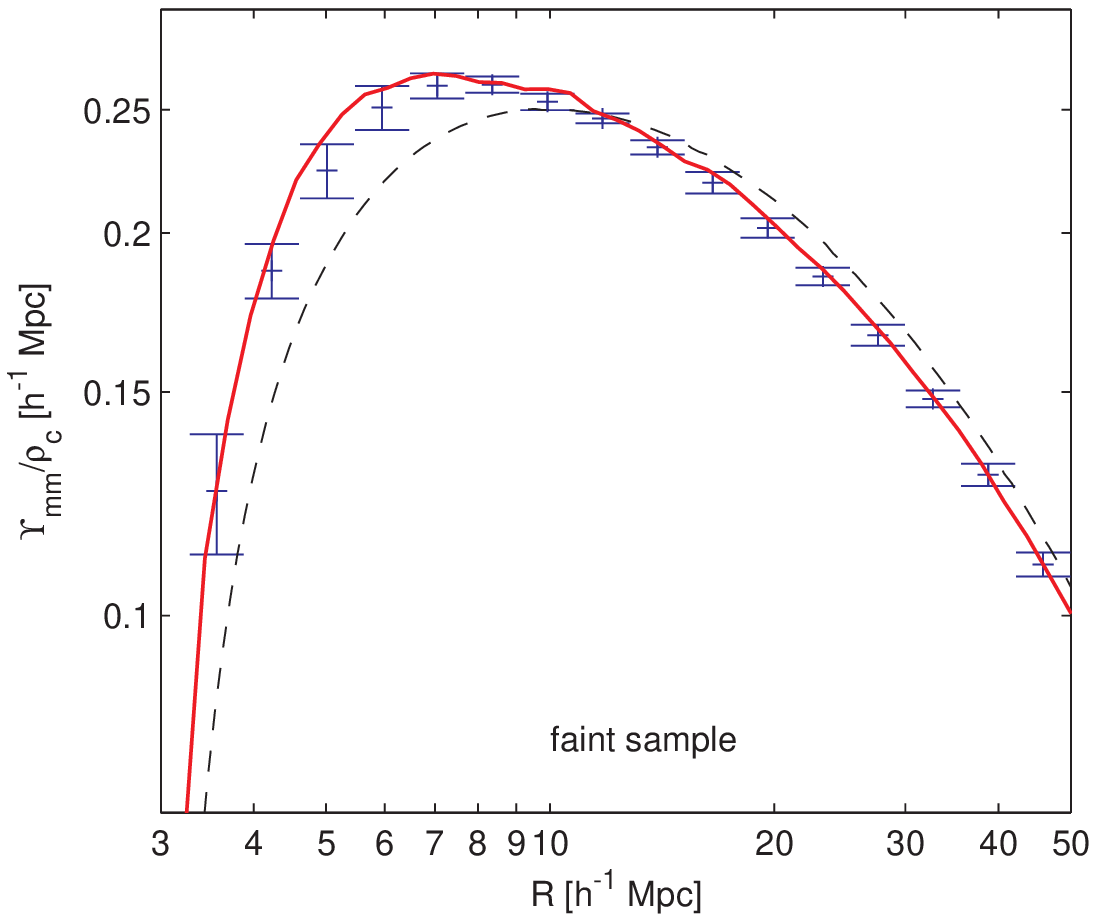}
	\caption{Reconstructed
	matter ADSD for the luminosity-threshold (left panel) and
	luminosity-bin sample (right panel). We plot the inferred
	value from the simulations including corrections for
	$r_\text{cc}\neq 1$ as well as the non-linear prediction
	derived from the measured matter-correlation of the
	simulations (red solid line) and the linear theory value (black
	dashed line). The non-linear correlation is well reproduced by the
	reconstruction, whereas there are remarkable deviations from
	linear theory on small scales.}
	\label{fig:reconfid}
\end{figure*}


\section{Numerical Results}

\label{sec:results}

To replicate the cross-correlation coefficient between galaxies and matter that
would be expected from real observational data, we use the artificial galaxy
catalogues described in the previous section to measure all of the statistics of
interest. Figure \ref{fig:rups} shows the resulting cross-correlation
coefficients of the luminosity-threshold (left panel) and
luminosity-bin (right panel) galaxy catalogues and two values of
$R_0$.  The ADSD is measured by counting the number of pairs in
cylinders with length $2\chi_\text{max}=100 \hMpc$ in real space. The errorbars
shown in these figures are derived from the standard-deviation between the eight
simulation volumes and thus represent the cosmic variance. We
again see that $\Delta\Sigma$ (black), corresponding to
$R_0=0$, leads to a cross-correlation coefficient that is strongly
scale dependent and different from unity, with 5--10\% deviations at
$10 \hMpc$. However, if we choose $R_0=3
\hMpc$, then we find that a cross-correlation coefficient close to
unity (blue with errorbars), with $r_\text{cc}=0.96$
at $4 \hMpc$, as predicted by perturbation theory for biased tracers
\cite{Seljak2009}. Furthermore, we observe this behaviour for both the
luminosity-bin sample and the luminosity-threshold sample. This
consistency suggests that the cross-correlation coefficient is largely
independent of the specific choice of the HOD used to generate the galaxy
catalogues, which is again consistent with the arguments in \cite{Seljak2009}
that the cross-correlation coefficient is nearly universal in the sense of being
only weakly dependent on the halo mass.
The theoretical prediction of Eq.~\eqref{eq:ccgamma} is plotted in
Figure \ref{fig:rups} as the blue solid line. For the latter we use
$\xi_\text{NL}$ to predict $\Upsilon_\text{mm}=\Upsilon_\xi$ and
$\Upsilon_\text{B}=\Upsilon_{\xi^2}$. We furthermore compare to the
full, non-expanded expression
\begin{equation}
 r_\text{cc}^{(\Upsilon)}(R)=\frac{\Upsilon_\xi+\alpha \Upsilon_\text{A}}
{\sqrt{\Upsilon_\xi(\Upsilon_\xi+2\alpha
\Upsilon_\text{A}+\alpha^2\Upsilon_\text{B}/2})}\label{eq:mucorrfull},
\end{equation} 
shown as the red-dash dotted line, whose range of validity is bounded
by the breakdown of perturbation theory rather than the relative
magnitude of the perturbation terms. Given the statistical
uncertainties of the direct simulation measurements, both expressions
are viable because the difference is $\sim2\%$ on the smallest scales
considered.

The results discussed above suggest, that we may invert Eq.~\eqref{eq:ccgam}
through the following,
\begin{equation}
	\Upsilon_\text{mm}(R)=\frac{\Upsilon_\text{gm}^2(R)}
{\Upsilon_\text{gg}(R)r_\text{cc}^2}\propto \Omega_\text{m}^2 \sigma_8^2
\label{eq:recovery}\ .
\end{equation}
The resulting statistic depends on the matter correlation
function and squared matter density, which is the usual parameter 
dependence of weak lensing measurements. This dependence enables us to constrain
cosmological
parameters.  Our theoretical model provides us with the scale
dependent correction factor $r_\text{cc}^{(\Upsilon)}$. 
Note that the cross-correlation coefficient is very close to unity on all scales
shown 
and even using $r_\text{cc}=1$ is acceptable given current observational
constraints. However, 
future observations will measure galaxy-galaxy lensing with much higher
statistical precision. The extraction of the full amount of
information contained in these measurements will require an
accurate modelling of $r_\text{cc}$. As argued in \cite{Seljak2009} this 
can be done in a relatively robust and model independent way. 

In Figure \ref{fig:reconfid}, we show the results of such a
reconstruction based on eight galaxy catalogues with the corresponding
cosmic variance errors. This reconstruction includes the
correction for the deviations of the cross-correlation coefficient from
unity. We see that the non-linear $\Upsilon_\text{mm}^\text{(nl)}$ is
reproduced with high accuracy, whereas the linear theory prediction
$\Upsilon_\text{mm}^\text{(lin)}$ deviates from our simulation
result. This finding is expected, because in Eq.~\eqref{eq:Upsilon} we subtract
$\Delta \Sigma(R_0)$ at $R_0=3\hMpc$, which is already at a non-linear scale. As
we go to larger scales this contribution is suppressed by $R_0^2/R^2$ 
and we slowly approach the linear theory predictions. 

\tr{Our numerical study implicitly uses the distant observer approximation, since we
project the density field in the simulation along one of the three Cartesian 
coordinate axes. In a real observation, the lines of sight to two nearby 
galaxies or to a foreground lens and a background source galaxy are inclined.
The question of whether the two statistics agree is related to the
extent to which the angular and the 2D projected power spectra agree. As 
discussed in \cite{LoVerde2008}, the angular power spectrum corresponds to
the 2D power spectrum if the Limber approximation \cite{Limber1953} is valid.
The LRG sample under consideration in our study has a median redshift of 
$z=0.23$ corresponding to a comoving distance $\chi_\text{l}=650 \hMpc$. 
Together with the maximum projection length $\chi_\text{max}=\pm 100 \hMpc$ and 
the maximum transverse distance to the galaxy $R_\text{max}=70\hMpc$, this 
corresponds to a maximum angle of $\theta_\text{max}=7.3^\circ$. The Limber 
approximation is typically precise to $<1\%$ for $l \approx \pi/\theta>10$ 
\cite{Smith2009b}, corresponding to $\theta<18^\circ$, and thus we can safely 
use the Cartesian analysis as an approximation for the observations.}

%

\section{Sources of Errors}\label{sec:rs}

Accurate studies of cosmological parameters require a careful
consideration of all effects that might change the
signal. In this section, we explore how large-scale redshift space
distortions and the difference between lensing and galaxy clustering
window functions impacts the reconstruction of the matter clustering.
\tr{Finally, we will discuss how strongly the radial bins are correlated.}

\subsection{Influence of Redshift Space Distortions}

\begin{figure*}[ht]
	\centering
	\includegraphics[width=1.0\textwidth]{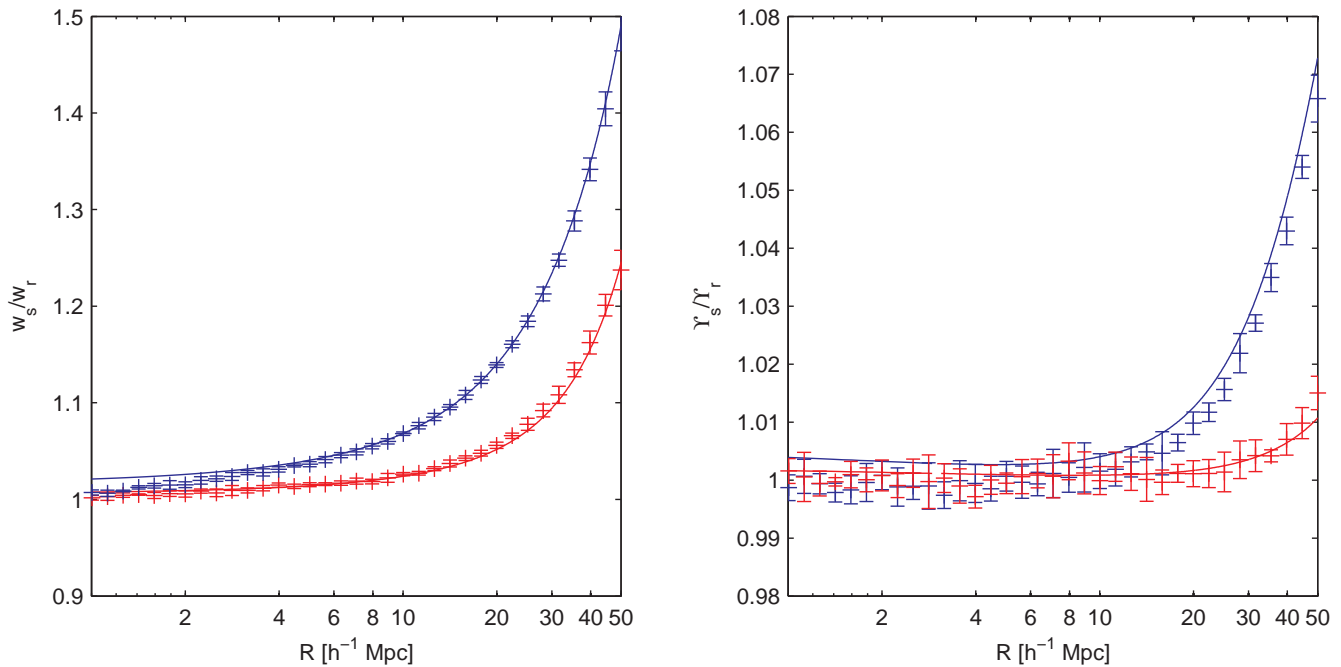}
	\caption{\emph{Left panel: }Residual effect of redshift space
	distortions on the projected galaxy-galaxy auto-correlation function
$w_\text{gg}$. We show the
	simulation measurements for the bright sample as crosses with
	errorbars for $\chi_\text{max}=50 \hMpc$ (upper blue) and
	$\chi_\text{max}=100 \hMpc$ (lower red) and the corresponding
	linear theory predictions. \emph{Right panel: }Residual effect
	on the annular differential surface density
	$\Upsilon_\text{gg}(R;R_0=3\hMpc)$ for the same integration
	lengths. As on the left panel, the upper blue curve and points
	correspond to $\chi_\text{max}=50 \hMpc$ whereas the lower red curve
	and points correspond to $\chi_\text{max}=100 \hMpc$.}
	\label{fig:rspro}
\end{figure*}

In large redshift surveys, such as the SDSS \cite{York2000} or 2dF
\cite{Colless2001}, the radial distance to a galaxy is inferred from
the recession velocity, under the assumption of a perfect Hubble law. In
reality, the coherent motions of galaxies and their virial
motions inside haloes will
add to the redshift and thus distort the inferred distance. In the
linear regime, on large scales these redshift space distortions can be
quantified using linear theory, neglecting virial motions within the
bound structures (for a review see \cite{Hamilton1998}). Following 
\cite{Kaiser1987} we can write the galaxy power spectrum
in redshift space in the plane parallel projection as
\begin{equation}
	P_\text{s}(k)=P_\text{r}(k) \left[1+\beta \mu
^2\right]^2\label{eq:psred}\ ,
\end{equation}
where $P_\text{r}(k)$ is the real space power spectrum of the tracer,
$\mu=\vec{k} \cdot \hat{\vec{x}}/k$ is the position angle with respect
to the redshift axis $\hat{\vec{x}}$ and $\beta=f(a)/b_1(a)$, where $f(a)\equiv
\derivd \ln D/\derivd\ln a$ is the logarithmic growth rate of fluctuations. We
obtain this directly by numerically evaluating the exact expression:
\begin{equation}
 f(a)=\frac{\derivd
\ln{H(a)}}{\derivd\ln{a}}+\frac{a}{(aH(a))^3}\frac{1}{\int_0^a \derivd a'
(a'H(a'))^{-3}}\ .
\end{equation} 

In what follows it will be convenient to rewrite Eq.~\eqref{eq:psred}
in terms of the Legendre polynomials $L_l(\mu)$
\begin{equation}
	P_\text{s}(k)=P_\text{r}(k) \left[\alpha_0 L_0(\mu)+\alpha_2
	L_2(\mu)+ \alpha_4 L_4(\mu)\right]\ ,
\end{equation}
where the coefficients are given by:
\begin{align}
\alpha_0(\beta)=&1+\frac{2}{3}\beta+\frac{1}{5}\beta^2\ ;\\
\alpha_2(\beta)=&\frac{4}{3}\beta+\frac{4}{7}\beta^2\ ;\\
\alpha_4(\beta)=&\frac{8}{35}\beta^2\ .
\end{align}
The redshift space correlation function is then obtained by a Fourier
transform of the power spectrum: 
\ba
\xi_{\rm gg,s}(r,\nu) & = & \frac{V}{(2\pi)^3}\int_0^\infty \derivd k k^2
P_\text{r}(k)
\int_{-1}^{1}\derivd \mu\left[1+\beta \mu^2\right]^2\nn\\
           &   & \times \int_0^{2\pi}
                  \derivd \varphi \eh{i\vec k\cdot \vec x}\label{eq:ftrs}\ ; \\
& = & \sum_{l=0}^{2}\alpha_{2l}(\beta)\xi_{2l}(r)L_{2l}(\nu),
\label{eq:xirs}
\ea
where $\nu$ is the angle between $\vec r$ and the axis along which the
redshift space distortion is present, i.e. $\nu=\hat{\vec x} \cdot
\hat{\vec r}=x/r$. The correlation function multipoles in the above
equation are defined as  
\begin{equation}
 \xi_{2l}(r)=(-1)^{l}\frac{V}{2 \pi^2}\int_0^\infty \derivd k k^2
P_\text{r}(k)j_{2l}(kr)\ ,
\end{equation} 
where the $j_l$ are the spherical Bessel functions:
$j_l(x)=J_{l+1/2}(x)/\sqrt{2x}$. We note that the above formulae are
equivalent to the formulation of \cite{Hamilton1992}. 

For our investigations, we are mainly concerned with the projected
correlation function or the closely related excess surface mass
density. The common assumption is that the integration along the line of
sight removes redshift space distortions. This assumption, however, would only
be correct in the limit of an infinite radial projection window, which is
not used in practice. 
Integrating Eq.~\eqref{eq:xirs} along the line of sight, we obtain
\ba
w_\text{gg,s}(R) & = &\int_{-\chi_\text{max}}^{\chi_\text{max}}\xi_{\rm
gg,s}(r,\nu)\derivd\chi \nn \\
                & = &
2\sum_{l=0}^{2}\alpha_{2l}(\beta)\int_0^{\chi_\text{max}}\!\!d\chi
\,\xi_{2l}\left(\sqrt{\chi^2+R^2}\right) \nn \\ 
& & \times \ L_{2l}\left(\frac{\chi}{\sqrt{\chi^2+R^2}}\right) \
.\label{eq:projectrs}
\ea
We shall use the above result to calculate the linear theory
predictions for the projected correlation functions in redshift space.
\begin{figure*}[ht]
	\centering
	\includegraphics[width=1.0\textwidth]{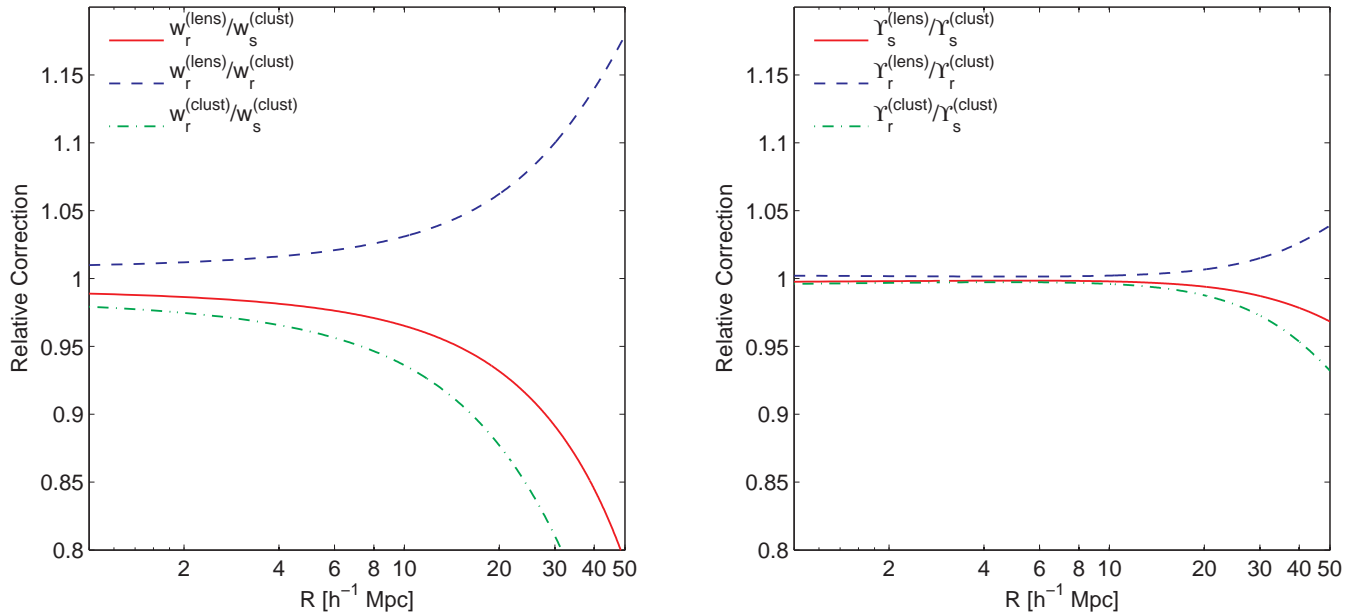}
	\caption{Correction factors that must be applied to the
          galaxy clustering measurements in order to remove redshift
          space distortions and to construct a quantity equivalent to
          the lensing signal. The total correction (red solid) is a
          product of a window correction (blue dashed) and a redshift
          correction (green dash-dotted). \emph{Left panel: }Projected
          galaxy-galaxy auto-correlation function $w_\text{gg}$. The
	  size of the corrections is quite remarkable on the largest scales.
	The apparent
	  effects on scales below $10 \hMpc$ arise from the flattening
	  of the linear power spectrum on these scales. Furthermore the
          correlation on these scales would be affected by the
          finger-of-god effects not included in our analysis.
	  \emph{Right panel: } Annular
          differential surface density for $R_0=3 \hMpc$. The residual
          correction is much smaller, $3 \%$ on the typical scales
          probed by galaxy-galaxy lensing. }
	\label{fig:ratiolenscorr}
\end{figure*}
In Figure \ref{fig:rspro}, we plot the ratio of the real to redshift
space projected correlation functions for the bright LRG galaxy sample with
$b=2.2$. We clearly see that the commonly used integration length of
$\chi_\text{max}=50 \hMpc$ leads to residual distortions of about
$40\%$ on scales $R\approx50 \hMpc$. These residual redshift space
  effects on the projected correlation function were previously 
discussed by \cite{Tinker2006,Tinker2007} (see also \cite{Padmanabhan2007}).
Moreover, we see that on these scales
the linear theory prediction is a very good description to the effects
that we observe in our simulations. The difference between linear
theory and simulation, on small scales, arises from the fact that we
do not model the virial motions, which cause the fingers-of-god. Furthermore,
the non-linear correlation function is more cuspy than the linear
correlation function on small scales. Therefore the linear predictions
in redshift space are boosted in amplitude by the compression
along the line of sight. The non-linear projected correlation function
is however much more influenced by the increased small-scale
clustering, and thus at small separations transverse to the line of
sight, it is less sensitive to the compression.  

Even though the linear prediction is a good description of the effect, 
removing it requires knowledge of the redshift space distortion parameter
$\beta$, 
which requires knowledge of both the cosmological model and bias. 
Since these are not known {\em a priory} but instead they must be determined
from the data. To do this accurately an iterative approach is needed, 
which complicates the analysis and ultimately limits the precision. Thus, 
it is advantageous if these corrections can be made as small as possible from
the onset. 
As we show in the right panel of Fig.~\ref{fig:rspro}, for the ADSD statistic
$\Upsilon$ much smaller residual corrections are required. 
The reduction is dramatic, with an order of magnitude smaller effect at the same
scale and for the same radial window. 
As discussed above, this reduction results from the compensated nature
of these statistics, which makes them much less sensitive to the long wavelength
fluctuations, so that the limit $\chi_\text{max} \rightarrow \infty$ is
approached faster. This makes these statistics
more attractive for practical applications than the projected correlation 
function $w$. In the context of galaxy clustering, 
similar compensated statistics have been proposed in \cite{Padmanabhan2007}.
Note that for lensing the typical radial window is hundreds of $\hMpc$ wide,
and for $\Delta \Sigma_\text{gm}$ or $\Upsilon_\text{gm}$ the effects of
redshift space distortions can be completely neglected on scales below $R\approx
100 \hMpc$. 

\tr{The analytical predictions for the impact of redshift space 
distortions on $\Upsilon$ and $w$ are based on the
Kaiser model and thus make use of the flat sky and distant observer
approximation. However, as we showed above, the ADSD is very robust to redshift 
space distortions with residual corrections on the $\mathcal{O}(1 \%)$ level. 
Due to the smallness of the correction and since our study is restricted to 
transverse separations that are small compared to the line of sight distance 
to the galaxies, our treatment is justified.}


\subsection{Dependence on projection length}


\begin{figure*}[ht]
	\centering
	\includegraphics[width=1.0\textwidth]{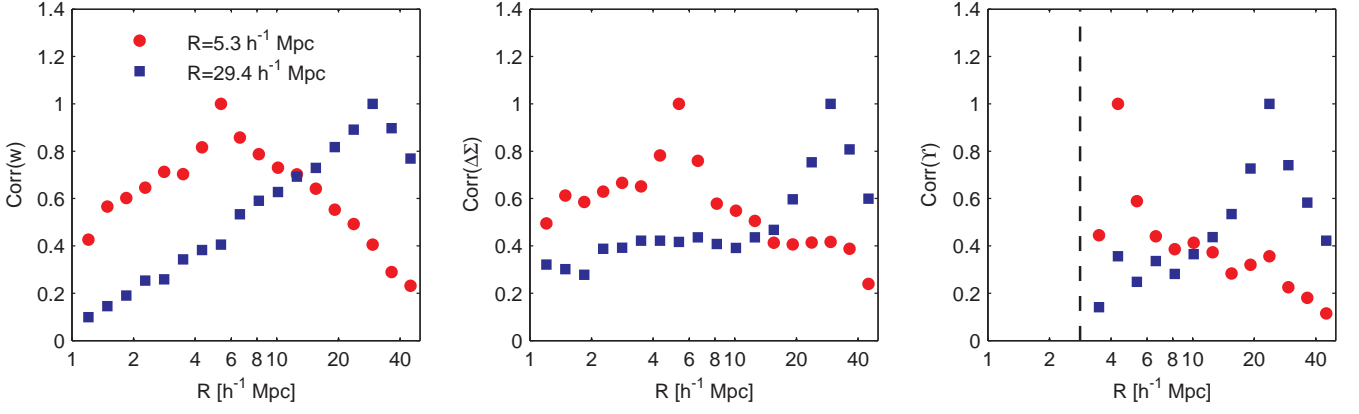}
	\caption{Columns extracted from the correlation matrix of the bright LRG sample
	\emph{Left panel: }$\text{Corr}(w)$ \emph{Central panel: }$\text{Corr}(\Delta\Sigma)$ \emph{Right panel: }
	$\text{Corr}(\Upsilon)$, where the vertical dashed line shows the cutoff radius $R_0$. Comparing the correlation matrices for $w$ and $\Delta\Sigma$ it is clear that
the width of the off-diagonal contributions is larger for the projected correlation function than for the excess
surface mass density. The same remains true if one compares the correlation of the ADSD and the projected correlation function.}
	\label{fig:covariance}
\end{figure*}

Our final goal is to compare a galaxy-matter cross-correlation
corresponding to a very broad lensing window and a galaxy-galaxy
auto-correlation that is calculated from a narrow top-hat window and
thus, in contrast to the lensing, subject to redshift space
distortions. In this context it is necessary to devise a correction
that accounts for both the redshift space distortions and the
different window functions. We already discussed the redshift space
effects and saw that they can be described by a scale dependent factor
$\Upsilon_\text{gg,r}^\text{(clust)}/\Upsilon_\text{gg,s}^\text{(clust)}$
given by linear theory. Here we use the superscript ``clust'' to
denote that this statistic is measured with the top-hat 
window. A similar numerical factor can be
used to transfer from the clustering to the lensing window
$\Upsilon_\text{gg,r}^\text{(lens)}/\Upsilon_\text{gg,r}^\text{(clust)}$.
The corrected galaxy correlation function corresponding to the lensing
measurements then reads as:
\begin{eqnarray}
\Upsilon_\text{gg,r}^\text{(lens)}&=&\Upsilon_\text{gg,s}^\text{(clust)}
\frac{\Upsilon_\text{gg,s}^\text{(lens)}}{\Upsilon_\text{gg,s}^\text{(clust)}}\
;\\
&=&\Upsilon_\text{gg,s}^\text{(clust)}
\underbrace{\frac{\Upsilon_\text{gg,r}^\text{(clust)}}
{\Upsilon_\text{gg,s}^\text{(clust)}}}_\text{redshift}
\underbrace{\frac{\Upsilon_\text{gg,r}^\text{(lens)}}
{\Upsilon_\text{gg,r}^\text{(clust)}}}_\text{integration length},
\end{eqnarray}
where $\Upsilon_\text{gg,s}^\text{(clust)}$ is the statistic that is
measured in the clustering survey and
$\Upsilon_\text{gg,r}^\text{(lens)}$ can be compared to
$\Upsilon_\text{gm,r}^\text{(lens)}$ measured from lensing.  

Figure \ref{fig:ratiolenscorr} shows the correction terms for the
bright LRG $(b=2.2)$ sample both for the projected correlation function $w$ and
the ADSD $\Upsilon$. The integration length correction is shown as a blue dashed
line.
The window correction was obtained from comparing the linear theory predictions
for $\Upsilon_\text{gg,r}^\text{(\text{lens})}$ and
$\Upsilon_\text{gg,r}^\text{(\text{clust})}$. Again we see that the
resulting corrections are much smaller for $\Upsilon$ than for $w$. This occurs
for the same reason 
as discussed above in the context of redshift space distortions: by using a
compensated 
window the sensitivity to long wavelength modes is removed and the limit 
$\chi_\text{max} \rightarrow \infty$ is approached faster, at which point the 
differences between different radial integration lengths disappear. 

Figure \ref{fig:ratiolenscorr} also shows 
the redshift factor as a green dash-dotted line and the final correction
as a solid red line. The redshift correction is the inverse of the
curve plotted in Fig.~\ref{fig:rspro}. 
We see that the projection length and redshift space effects go in the opposite
direction. This partial cancellation further minimises their effect, so that for
the ADSD $\Upsilon$, their combined effect is less than $3\%$ even at
$R \approx 50 \hMpc$.  

\subsection{Covariance matrix}
\tr{Another benefit of the compensated ADSD is that its correlation matrix
$\text{Corr}_{ij}=\left\langle \Upsilon_i\Upsilon_j \right\rangle/\sqrt{\left\langle \Upsilon_i\Upsilon_i \right\rangle \left\langle \Upsilon_j\Upsilon_j \right\rangle}$
has weaker off-diagonal contributions than that for the projected correlation 
function. Usually two point statistics such as the correlation function
show strong correlations between different radial bins, i.\,e. important 
off-diagonal entries in the covariance matrix. The compensated window 
Eq.~\eqref{eq:upswindow} relating the ADSD to the correlation function reduces 
these off-diagonal contributions remarkably. This statement refers to the
cosmic variance contribution to the covariance matrix only. The shape noise
adds predominantly to the diagonal error and thus further reduces the 
off-diagonals of the correlation matrix.}
\par
\tr{In Figure \ref{fig:covariance} we show columns extracted from the correlation 
matrices of $w$, $\Delta\Sigma$ and $\Upsilon$, respectively, for the bright LRG
sample. 
The covariance matrix is estimated by calculating the variance over $160$ 
subvolumes of $750 \times 750 \times 300\ h^{-3}\text{Mpc}^3$. From this plot 
it is obvious that the off-diagonal contributions to the covariance matrix are 
reduced as one transitions from the projected correlation to the excess surface 
mass density and ADSD. One would expect some additional covariance due to the 
subtraction of $\Delta\Sigma(R_0)$ in $\Upsilon$, but it turns out that the 
reduced off-diagonal covariance remains for the ADSD. We compare the 
signal-to-noise $(S/N)^2_{\Delta\Sigma}=\sum_{i,j}\Delta\Sigma_{i}C^{-1}_{ij}\Delta\Sigma_j$ 
and $(S/N)^2_\Upsilon=\sum_{i,j}\Upsilon_{i}C^{-1}_{ij}\Upsilon_j$, where $C$ are 
the covariance matrices of $\Delta\Sigma$ and $\Upsilon$, respectively.
The sum runs over radial bins with $R_i>1 \hMpc$  for $\Delta\Sigma$ and 
$R_i>R_0$ for $\Upsilon$.
We see that the signal-to-noise ratio is degraded by a factor 
$0.38 < (S/N)_\Upsilon/(S/N)_{\Delta\Sigma} < 0.45$
over a range of cutoff radii $5 \hMpc > R_0 > 1 \hMpc$. So the advantage of being
able to interpret the result in terms of perturbation theory just has to be paid
by a factor $2-3$ decrease in signal-to-noise.}



\section{Variant Cosmologies}\label{sec:varcosm}

\begin{figure*}[!p]
	\centering
	\includegraphics[width=1.0\textwidth]{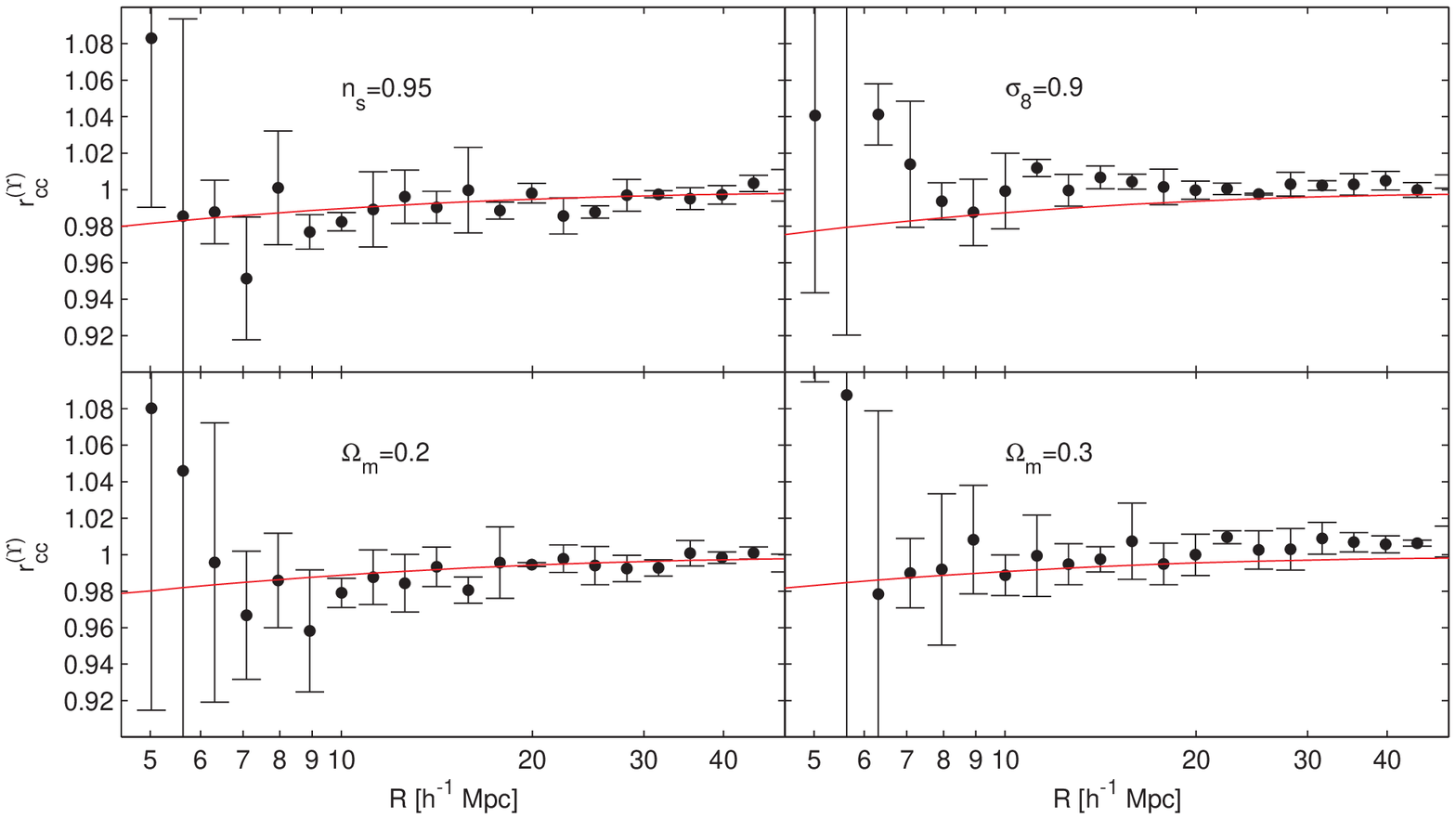}
	\caption{Cross-correlation coefficient of the ADSD $\Upsilon$
          for the full galaxy sample. The panels show simulation
          output as circles with errorbars and theoretical predictions of
Eq.~\eqref{eq:ccgamma}
          for $R_0=5 \hMpc$ (red solid lines).
	\emph{Top left panel: }Reduced spectral index $n_\text{s}=0.95$ 
	\emph{Top right panel: }Increased normalisation $\sigma_8=0.9$
	\emph{Bottom left panel: }Reduced matter density $\Omega_\text{m}=0.2,\
\Omega_\Lambda=0.8$
	\emph{Bottom right panel: }Increased matter density
$\Omega_\text{m}=0.3,\ \Omega_\Lambda=0.7$ We see that the increased number of
high mass haloes in the
        $\Omega_\text{m}=0.3$ and $\sigma_8=0.9$ models leads to a
        higher number of satellite galaxies and thus partially
        compensates the drop of the cross-correlation coefficient for
        haloes on small scales.}
	\label{fig:ccvarcosm}
	\includegraphics[width=1.0\textwidth]{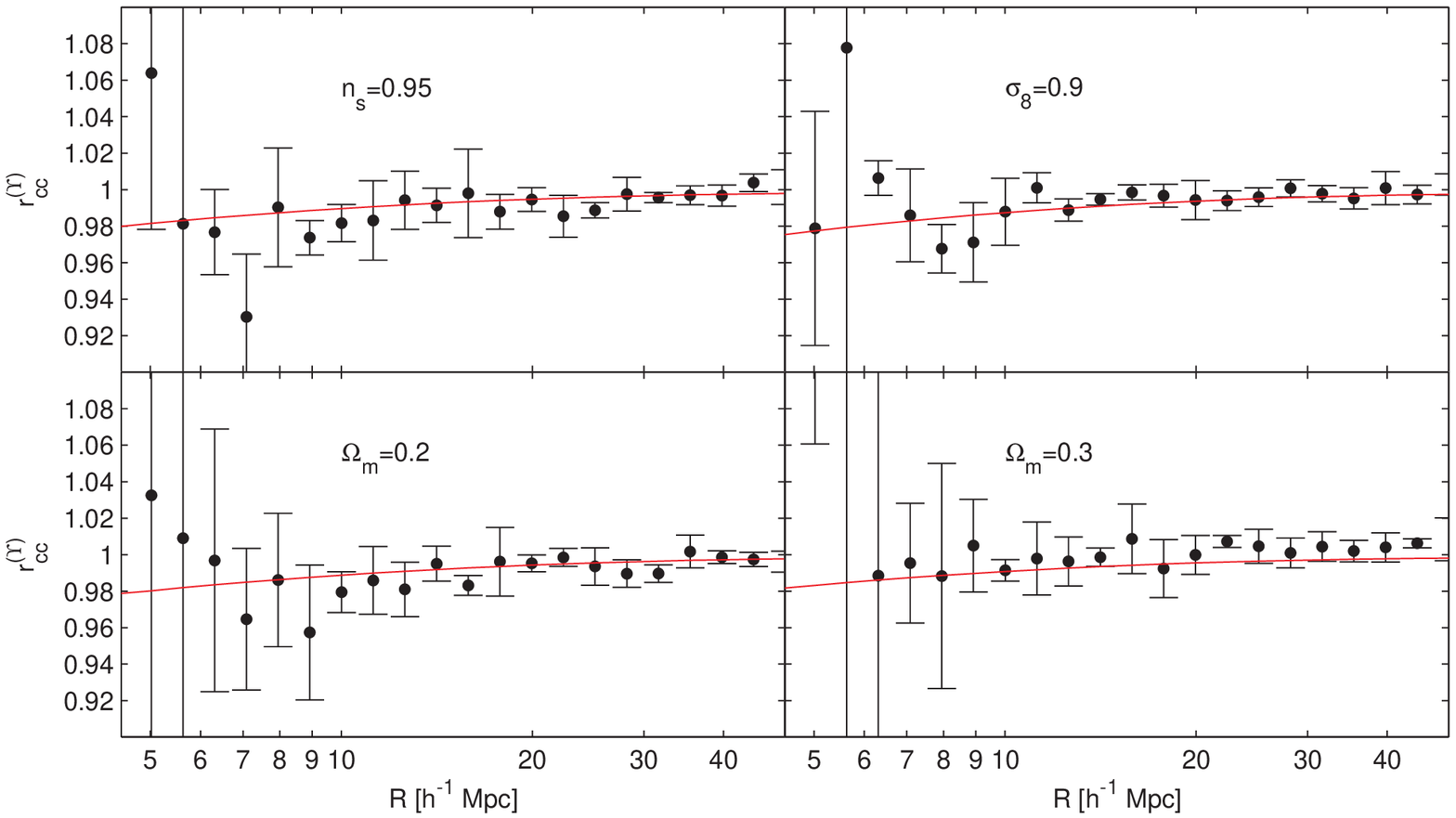}
	\caption{Same as Fig. \ref{fig:ccvarcosm} but for the central LRGs. The
central LRGs are a cleaner representation of a halo sample and thus better
reproduce our theoretical expectations.}
	\label{fig:ccvarcosmcent}
\end{figure*}


\begin{figure*}[!t]
	\centering
	\includegraphics[width=1.0\textwidth]{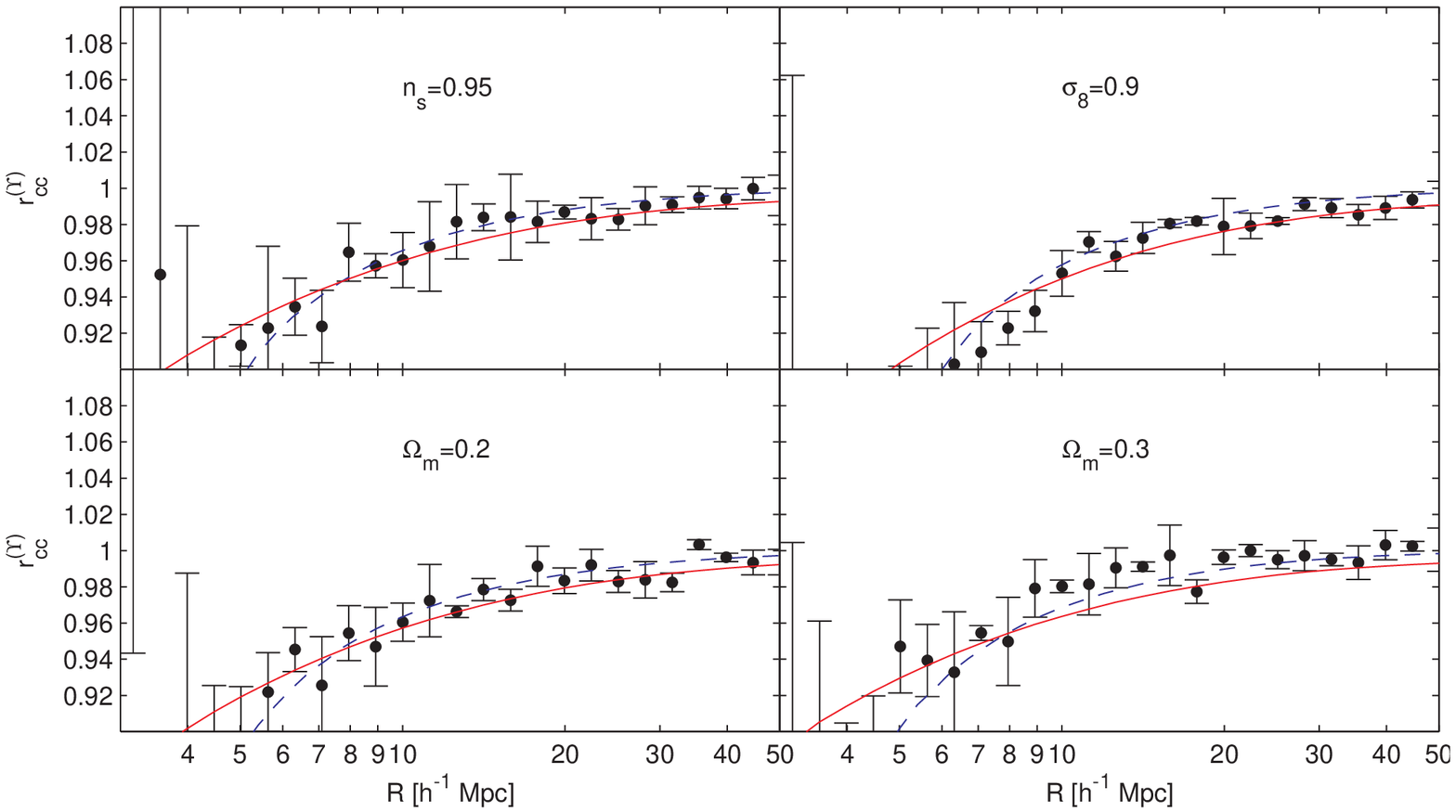}
	\caption{Same as Fig. \ref{fig:ccvarcosm}, but for $R_0=3 \hMpc$ and for
LRG sample from which the clusters with mass exceeding $M=3\tim{14} \hMs$ and
all satellites were removed. Due to the cluster subtraction the bias ratio
changes to $\alpha=0.41$. In addition to the usual correction (red solid line)
we also plot the phenomenological correction $r_\text{cc}\approx 1-\alpha^2
\xi(R/2)/4$ (blue dashed line) The cluster subtraction also allowed us to reduce
the cutoff radius to $R_0=3 \hMpc$.}
	\label{fig:ccvarcosmclust}
\end{figure*}

In the previous sections, we presented results for our mass clustering
reconstruction for one specific cosmological model. In this section,
we explore how well the $\Upsilon_\text{mm}$ reconstruction performs
for four variations to our fiducial model.

The four variations are presented in Table~\ref{tab:cosmoparam}, and
we denote them by C1-C4. Each of these models differs from the
fiducial model in exactly one parameter, and for each variation we
have performed four simulations providing a volume of $V=13.5 \
h^{-3}\unit{Gpc}^{3}$. We populate these simulations using the same HOD
parameters inferred for the luminosity-threshold LRG sample as
described in \S\ref{sec:num}.

Figure~\ref{fig:ccvarcosm} shows the cross-correlation coefficient
inferred from the statistic $\Upsilon$, for $R_0=5 \hMpc$ for the full
LRG sample. The reason for increasing the cutoff radius is that the cluster
masses for some of the variant cosmologies are increased and thus we apply this
more conservative cutoff radius to ensure $r_\text{cc}\approx1$. We then see
that the variation of cosmology does not
significantly change the general trends observed for our fiducial
model.  We notice that there is a weak scale dependence, and in all
cases the trend to lower $r_\text{cc}$ is well described by the theoretical
model given by Eq.~\eqref{eq:mucorrfull}, which is over-plotted as a
green line. There appears to be some small $(\sim5\%)$ discrepancy for
the higher $\sigma_8$ model at smaller scales $R<10 \hMpc$, where we see
an increase in the cross-correlation coefficient.  This is likely due to
the fact that the cut-off scale $R_0$, is actually less than twice the
virial radius of the most massive haloes, and so the statistics are
still sensitive to the internal structure of the haloes.

We examined whether the agreement might be further improved through
using only the central LRG galaxies in the reconstruction.  The
results of this test are presented in Figure~\ref{fig:ccvarcosmcent},
and we indeed find better agreement. We believe that this is due to the fact 
that the influence of satellite-satellite pairs 
from massive clusters has been removed. This means that the central
galaxy sample is closer to a mass-selected halo sample and is thus
less influenced by the details of how galaxies populate haloes.

A simple way to further reduce this sensitivity is to eliminate the most
massive haloes
from the data. Since these contain many galaxies, they are easy to identify
in an observation. In Figure \ref{fig:ccvarcosmclust}, we plot the
cross-correlation coefficient of a halo sample, from which we removed all the
clusters with mass exceeding $M\ge 3 \tim{14} \hMs$ and all the central
galaxies. Having removed the clusters, we can lower the cutoff radius to $R_0=3
\hMpc$. Clearly the cross-correlation coefficient shows stronger deviations from
unity. These are, however, better reproduced by our model than the full or
central sample. To achieve this agreement we needed to change the bias ratio
$\alpha$ accounting for the new upper mass threshold. The corrections in Figure
\ref{fig:ccvarcosmclust} use $\alpha=0.41$ instead of our fiducial choice of
$\alpha=0.26$.

\begin{figure}[ht]
	\centering
	\includegraphics[width=0.49\textwidth]{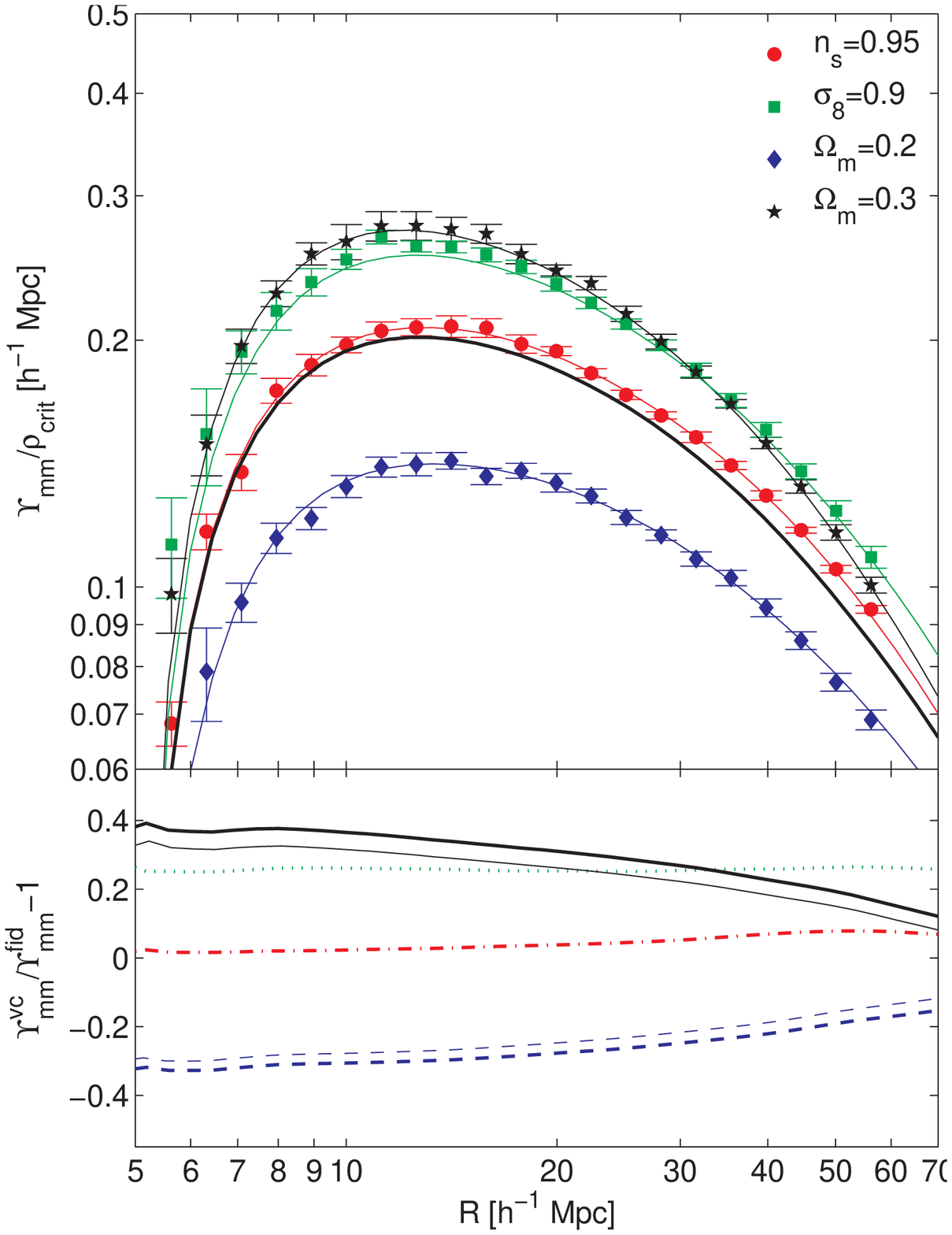}
	\caption{\emph{Top panel: }Reconstructed matter ADSD of the
          variant cosmologies for $R_0=5 \hMpc$. The points with
          errorbars show the simulation results for the four variant
          cosmologies as measured in the simulations, whereas the solid
          lines show the corresponding non-linear matter correlation
          function. From top to bottom: $\Omega_\text{m}=0.3$ (black
          stars), $\sigma_8=0.9$ (green squares), $n_\text{s}=0.95$
          (red circles) and $\Omega_\text{m}=0.2$ (blue diamonds). The
          thick black line is the non-linear matter correlation
          function of our fiducial model plotted here for reference.
          \emph{Bottom panel: }Fractional difference of the
          reconstructed matter statistics with respect to the fiducial
          model. From top to bottom we show $\Omega_\text{m}=0.3$ (solid black),
          $\sigma_8=0.9$ (green dotted), $n_\text{s}=0.95$ (red dash dotted) and
          $\Omega_\text{m}=0.2$ (blue dashed). For the $\Omega_\text{m}=0.2$ and
          $\Omega_\text{m}=0.3$ cosmologies we also include the effect
          of wrong \emph{a priori} cosmology as thin lines with corresponding line style (for
          further discussion see text). }
	\label{fig:reconvar}
\end{figure}

Another way to improve agreement between theory and simulation is to use a
phenomenological correction factor. Inspired by the fact that the correction for
the cross correlation coefficient $r_\text{cc}^{(\xi)}$ in
Equation~\eqref{eq:ccxitheo} is proportional to the correlation function, we can
simplify our correction factor using the approximation
$r_\text{cc}^{(\Upsilon)}(R)\approx1-\alpha^2\xi(R/2)$/4. The argument $R/2$ in
the correlation function can be motivated considering the window for the
correlation function plotted in the right panel of Figure \ref{fig:PWindow}.
There we see that the windows for $\xi(R/2)$ and $\Upsilon(R)$ peak at
approximately the same scale in k-space. We over-plot the phenomenological
correction as the red dashed line in Figure \ref{fig:ccvarcosmclust}. With this
replacement we can slightly improve the agreement between theory and
measurement.

In Figure \ref{fig:reconvar} we reconstruct the matter correlation
$\Upsilon_\text{mm}(R)$ from the simulation measurements of
$\Upsilon_\text{gm}(R)$ and $\Upsilon_\text{gg}(R)$ as the points with
errorbars. For this plot we use the full galaxy samples, whose cross-correlation
coefficient was shown in Figure \ref{fig:ccvarcosm}. We see that the non-linear
matter correlation function is
reproduced for all the four variant cosmologies. Furthermore, there are
clear differences both in shape and amplitude between the different
cosmologies, so that inference of cosmological parameters should be
feasible. Differences in $\Omega_\text{m}$ and $\sigma_8$ are more
prominent than the effect of changing the slope of the primordial
spectrum $n_\text{s}$. The small discrepancy between the simulations and the
theoretical prediction
for the high $\sigma_8$ model in Fig.~\ref{fig:ccvarcosm} translates
into a tension between inferred and real matter ADSD.

The lower panel of Figure~\ref{fig:reconvar} emphasises the possibility of
inferring cosmological parameters by showing the fractional
differences in the recovered ADSD for the different cosmologies C1 -- C4 with
respect to the fiducial model. Variations in the slope of the power spectrum
differ from the fiducial model only on the $5 \%$ level at $R=30 \hMpc$, whereas
the quadratic dependence of the estimator on $\sigma_8 \Omega_\text{m}$ leads to
a clear separation of the variant $\sigma_8$ and $\Omega_\text{m}$ models from
the fiducial model ($25 \%$ at $R=30 \hMpc$). If the lensing study extends to
sufficiently large scales $\sigma_8$ and $\Omega_\text{m}$ are separable by
their shape. Here we are using the fact that a change in $\Omega_m$ or
$\sigma_8$ affects the amplitude as well as the shape of the correlation
function.

One caveat is that the inference of $\Upsilon_\text{mm}$ requires the assumption
of an \emph{a priori} cosmology. This
assumption enters the reconstruction in three places. Firstly, we are
using the clustering and lensing measurements as a function of the
distance transverse to the line of sight. The observation, however,
provides both clustering and tangential shear distortions as a function of
angular separation. To relate the two, one needs to calculate the
angular diameter distance to the foreground galaxy sample, which
depends on $\Omega_\text{m}$. A wrong prior on the cosmological model would thus
cause
a horizontal shift in the inferred statistic. Secondly, the definition
of the excess surface mass density includes the critical surface mass
density, a ratio of the angular diameter distances to the lens, the
source and between the two. The latter affects the amplitude of
$\Upsilon_\text{mm}$ in quadrature. Thirdly, we use cosmology to 
compute the cross-correlation coefficient (Eq.~\ref{eq:ccgamma}). 
This also only has a weak dependence on 
cosmology, since the cross-correlation coefficient is close to unity to start
with.

In order to estimate the magnitude of the first two effects we pose the
following
question: How is the inferred statistic for $\Omega_\text{m}=0.2$ or
$\Omega_\text{m}=0.3$ affected if we wrongly assume the fiducial
cosmology, $\Omega_\text{m}=0.25$, for the measurement? As a
reasonable example, we take $z_\text{l}=0.25$ and $z_\text{s}=0.5$, for
the lens and source redshifts. For these cases, we obtain a $2\%$
increase (decrease) in $\Sigma_\text{crit}$ for $\Omega_\text{m}=0.2$
($\Omega_\text{m}=0.3$) with respect to the fiducial
$\Omega_\text{m}=0.25$. These results are shown as the thin lines in
the lower panel of Fig.~\ref{fig:reconvar}.  The shift caused by the
cosmology dependence of the angular diameter distance to the lens
galaxy has a smaller effect and is on the order of $\sim1\%$.
One route to remove part of this dependence from the measurement is to change
the estimator
$\Upsilon_\text{mm}\to\Upsilon_\text{mm}/\Sigma_\text{crit}^2$. This
can be done by writing
$\Upsilon_\text{gm}=\Sigma_\text{crit}\left[\overline{\gamma}_\text{t}
(R)-R_0^2/R^2\gamma_\text{t}(R_0)\right]$
and substituting this expression into Eq.~\eqref{eq:recovery}
\begin{equation}
 	\frac{\Upsilon_\text{mm}(R)}{\Sigma_\text{crit}(\Omega_\text{m})^2}=
\frac{\left[\overline{\gamma}_\text{t}(R)-R_0^2/R^2\gamma_\text{t}(R_0)\right]^2
}
{\Upsilon_\text{gg}(R)r_\text{cc}^2}\ .
\end{equation}
The benefit of this redefinition is that the quantity we compare to theory has
one cosmology dependence less, and $\Sigma_\text{crit}$ can be calculated for
each tested cosmological model. However, the angular diameter distance still
depends on $\Omega_\text{m}$. We could introduce another factor that takes care
of 
this dependence, but  for SDSS data at low redshift the effect is small. In
general one can use  
an iterative procedure or check whether within the errors 
on $\Omega_\text{m}$ the effects exceed observational errors. A similar
iterative procedure  
can be used to verify the sensitivity to the assumed value of the
cross-correlation coefficient 
in the reconstruction.

\subsection*{Reconstruction Procedure}

To conclude, we summarise our procedure for inferring the matter
clustering from lensing and clustering measurements in terms of the following
five steps:

\begin{enumerate}
\item Measure galaxy-galaxy lensing signal $\gamma_\text{t}$
  for a certain lens galaxy sample, and calculate
  $\Delta\Sigma_\text{gm}(R)$ from the tangential shear. This
  first step requires the assumption of an \emph{a priori}
  cosmological model that has to be confirmed or refuted by the
  final result.
\item Measure the galaxy-galaxy clustering of the lens galaxy
  sample and calculate the projected correlation function
  $w_\text{gg}(R)$. Integrate the result to obtain
  $\Delta\Sigma_\text{gg}(R)$.
\item Estimate the typical host halo virial radius of the galaxy sample
  under consideration. Use this estimated $R_0$ to correct for the
  central contributions in $\Delta\Sigma_\text{AB}(R)$ by calculating
  $\Upsilon_\text{AB}(R)=\Delta\Sigma_\text{AB}(R)-\Delta\Sigma_\text{AB}(R_0)R_0^2/R^2$.
\item Make predictions for the transfer function and resulting
  matter auto-correlation functions for a set of cosmological
  parameters and/or modifications of gravity. Use these to
  calculate $\Upsilon_\text{mm}^{(\text{theo})}$ and find the
  best fit parameters by comparison to the empirical result.
\item Iterate until convergence.
\end{enumerate}

\section{Conclusions} \label{sec:discuss}

In our study, we examined how well can one reconstruct the 
dark matter clustering from observations of galaxy clustering combined with galaxy-galaxy lensing. This reconstruction procedure could for instance be applied to the SDSS galaxy survey, in particular the Luminous Red Galaxies. 
In a first step, we generated realistic LRG galaxy catalogues for both
a luminosity-threshold and a luminosity bin sub-sample of the LRGs.
We then used these galaxy catalogues to extract information about the
cross-correlation coefficient between galaxies and matter.

We introduced a new statistic $\Upsilon(R)$, which we termed the
Annular Differential Surface Density (ADSD), that removes the
influence of small, non-linear scales on the excess surface mass
density. This subtraction is necessary since the scales smaller than
the virial radius of the haloes are dominated by the halo profile
rather than the pre-shell-crossing evolution of the large scale
cosmological fluid. Both numerical studies and theoretical
calculations indicate that the cross-correlation coefficient of the
ADSD is close to unity and that the residual scale dependence is well
described by an analytic correction.  Having focused our
investigations on the excess surface mass density
$\Delta\Sigma_\text{gm}$, our results can be directly applied to
measurements of galaxy-galaxy lensing and the projected galaxy
correlation function.

We also studied systematic effects that might bias the comparison of
lensing and galaxy clustering measurements. In terms of the projected
correlation function, both numerical studies and a linear theory
treatment, following \cite{Kaiser1987}, show that the common integration
over $\pm 50\hMpc$ along the line of sight is still biased by redshift
space distortions. We have shown the necessity of correcting for the
large scale peculiar motions in any such clustering
measurement. We also investigated the effect of different window functions used 
for lensing and clustering. These introduce additional effects, that must be 
accounted for in the final analysis. We have found that the ADSD statistic
$\Upsilon(R)$ is much less sensitive to both of the effects, and to long 
wavelength modes in general, than the usual projected correlation function $w(R)$,
because of the (partially) compensated nature of its transverse window. 

As our key result, we devised a method to recover the dark matter
clustering from galaxy-galaxy lensing and galaxy clustering
measurements using the cross-correlation coefficient for the
ADSD. Assuming $r_\text{cc}=1$ for simplicity leads to at most $8\%$
bias on scales below $R\approx 5 \hMpc$ in the recovered statistic
$\Upsilon_\text{mm}(R)$. We can however remove this bias based on our
theoretical modelling of the scale dependence of the cross-correlation
coefficient. The main advantage of our method is that the galaxy dependence is
scaled out of the equations, since the theoretical model predictions for the
cross-correlation coefficient between haloes and dark matter are relatively
independent of the halo mass over a wide range of mass. Thus, we believe that
the method devised here is more robust than the methods which are based on HOD
fitting (e.\,g.\ \cite{Yoo2009}), which fit for the cosmological parameters and
the HOD parameters jointly, marginalising over the uncertainties in the HOD
parameters.

A study on four other cosmological models verified the robustness of the new
estimator. Varying one parameter of the fiducial $\Lambda$CDM model at a time,
we found that the cross-correlation coefficient shows a scale dependence
consistent with the fiducial model. We were able to reconstruct the ADSD of the
matter correlation function, and the inferred statistic $\Upsilon_\text{mm}$ can
be used to distinguish cosmological models both, from the shape and the
amplitude of the recovered statistic. This study also showed that, if one is
capable of accurately distinguishing central from satellite galaxies and/or
remove clusters, then one can eliminate the influence of satellite galaxies and
so render the cross-correlation coefficient closer to theoretical predictions
for haloes in numerical simulations.  These advantages make it worthwhile to
define a clean central galaxy sample and to remove the clusters. While the
non-linear matter correlation can be recovered with high fidelity, the linear
correlation is only recovered at large scales. This fact strengthens the need
for a well developed and tested perturbation theory of large-scale clustering
that extends into the weakly non-linear regime and which can thus provide us
with an estimator of $\xi_\text{NL}$ without having to carry out simulations for
each cosmological model. 

The ADSD statistic subtracts out a lensing signal at $R_0$
(Eq.~\ref{eq:Upsilon}). This subtraction procedure decreases the signal-to-noise
on the inferred statistic $\Upsilon$ as compared to $\Delta\Sigma$. This price
seems worth paying, since it brings the cross-correlation coefficient much
closer to unity with residual deviations from unity that are well understood
theoretically. An application of this method to observational data will have to
address the problem of estimating $\Delta\Sigma(R_0)$. The cubic spline fit used
for our numerical studies will not be appropriate given the large statistical
fluctuations in observed lensing signal. Several alternatives to estimate
$\Delta\Sigma(R_0)$ are explored in \cite{Mandelbaum2009}, with the most
successful being a fit with a running power-law (three parameters) to the radial
bins around $R_0$.

Our numerical results are based on the SDSS spectroscopic LRG sample, i.\,e.\ on
the galaxies living in the most massive haloes. Based on the success and
generality of the theoretical model we expect that a similar
behaviour for the Main spectroscopic sample galaxies in the SDSS, which
live predominantly in lower mass haloes. The lower halo masses may
also enable a lower cutoff radius $R_0$, especially if haloes with higher 
mass are effectively removed from the sample. \tr{If the halo sample spans a 
wide range of masses, it should be split 
into mass bins and $R_0$ should be chosen appropriately for each of the mass 
bins.} We shall reserve this topic for future investigation.


\begin{acknowledgements}
We acknowledge Vincent Desjacques, Patrick McDonald and Jeremy Tinker for helpful discussions; Volker Springel for making public \texttt{GADGET-II} and for
providing his \text{B-FoF} halo finder; Roman Scoccimarro for making
public his \texttt{2LPT} code. T.\,B. gratefully acknowledges support by
a grant of the German National Academic Foundation during the initial phase of
this project. R.\,E.\,S. is supported by a Marie Curie Reintegration
Grant. R.\,M. was supported for the duration of this work by NASA through Hubble
Fellowship grant \#HST-HF-01199.02-A awarded by the Space Telescope Science
Institute, which is operated by the Association of Universities for Research in
Astronomy, Inc., for NASA, under contract NAS 5-26555. This work is partly
supported by the Swiss National Foundation under contract 200021-116696/1,
Packard Foundation and WCU grant R32-2008-000-10130-0.
\end{acknowledgements}

\bibliographystyle{../arxiv_physrev}
\bibliography{paper}

\end{document}